%% file: contourmaps.tex
\documentclass[a4paper,12pt]{article}

\usepackage[scale={0.7,0.9},centering,includeheadfoot]{geometry}

\input{contourmaps_style.tex}

\title{\bf{Quantifying the uncertainty of \\ contour maps}}
\author{\scshape{David Bolin$^{a}$ and Finn Lindgren$^{b}$}\\
{\small $^{a}$Mathematical Sciences, Chalmers and University of Gothenburg}\\
{\small $^{b}$Mathematical Sciences, University of Bath}}
\date{}
\begin{document}
\maketitle
\begin{center}
\begin{minipage}{0.9\textwidth}
  \noindent{\bf Abstract:}
  \input{contourmaps_abstract.tex}

\vspace{0.3cm}\noindent{\bf Key words:}
Excursions; Kriging; Latent Gaussian processes; Markov random fields;
Spatial uncertainty visualization
\end{minipage}
\end{center}

\input{contourmaps_text.tex}

\bibliographystyle{plainnat}
\bibliography{journ_abrv,completebib,contourmapadd}


\end{document}

%% file: contourmaps_style.tex
\usepackage{amssymb,amsmath,amsfonts,amsthm}
\usepackage{natbib}
\usepackage[latin1]{inputenc} 
\usepackage[T1]{fontenc}
\usepackage{verbatim}
\usepackage{graphicx}
\usepackage{booktabs}
\usepackage{color}
\usepackage{pbox}

\DeclareMathOperator{\R}{\mathbb{R}}

\DeclareMathOperator*{\argmax}{arg\,max}

\newtheorem{thm}{Theorem}[section]
\newtheorem{defn}[thm]{Definition}

\newcommand{\proper}{\mathsf}
\newcommand{\pP}{\proper{P}}
\newcommand{\pE}{\proper{E}}
\newcommand{\pV}{\proper{V}}

\newcommand{\pN}{\proper{N}}
\newcommand{\mv}[1]{{\boldsymbol{\mathrm{#1}}}}
\newcommand{\md}{\ensuremath{\,\mathrm{d}}}
\newcommand{\exset}[2]{E_{{#1}}^{{#2}}}
\newcommand{\mapset}{G}

\newcommand{\s}{{\mathbf{s}}}
\newcommand{\uu}{{\mathbf{u}}}

\newcommand{\mat}[1]{\begin{bmatrix}#1\end{bmatrix}}

\graphicspath{{figs/}}

%% file: contourmaps_abstract.tex
Contour maps are widely used to display estimates of spatial
fields. Instead of showing the estimated field, a contour map only
shows a fixed number of contour lines for different levels. However,
despite the ubiquitous use of these maps, the uncertainty associated
with them has been given a surprisingly small amount of attention. We
derive measures of the statistical uncertainty, or quality, of contour
maps, and use these to decide an appropriate number of contour lines,
that relates to the uncertainty in the estimated spatial field. For
practical use in geostatistics and medical imaging, computational
methods are constructed, that can be applied to Gaussian Markov random
fields, and in particular be used in combination with integrated
nested Laplace approximations for latent Gaussian models. The methods
are demonstrated on simulated data and an application to temperature
estimation is presented.

%% file: contourmaps_text.tex
\section{Introduction}
\begin{figure}[t]
\begin{center}
\resizebox{0.8\linewidth}{!}{\includegraphics{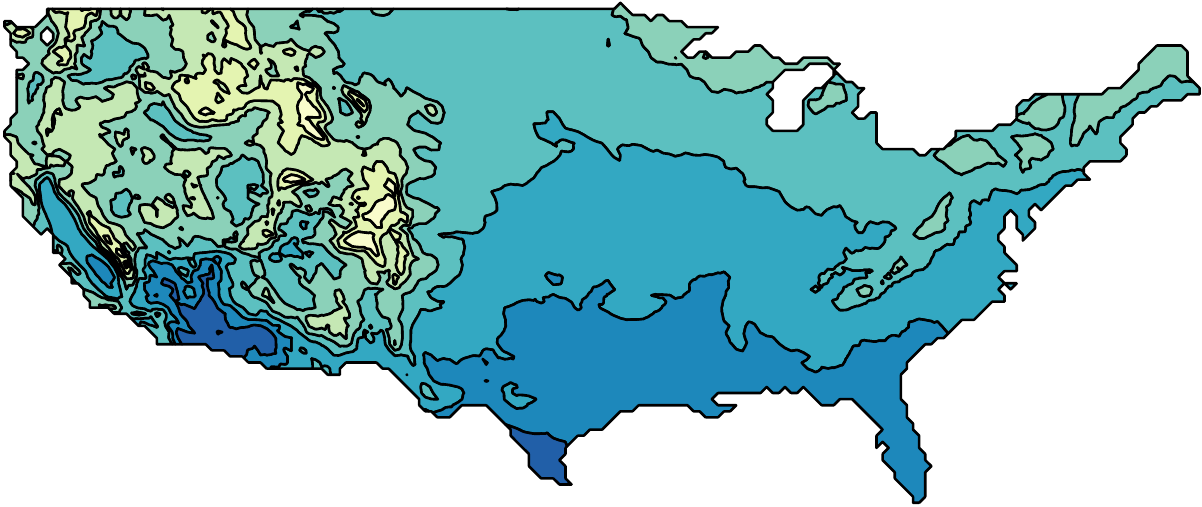}}%
\resizebox{0.07\linewidth}{!}{\includegraphics{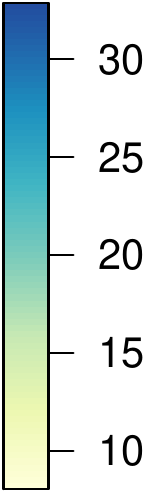}}
\end{center}
\caption{Example of a contour map for the mean summer temperature in 1997 in the US. The temperature changes due to topography in the western part is apparent. The challenge is to quantify our level of trust in the level of detail in the contour curves.}
\label{fig:temp_cm1}
\end{figure}

Contour maps are often used in environmental statistics and applied spatial statistics to display estimates of continuous surfaces. One such example can be seen in Figure~\ref{fig:temp_cm1} where an estimate of the mean summer temperature in 1997 is shown for the US.


One of the earliest documented uses of contour maps is from 1548, when water depth was displayed using contours in a map \citep{thrower2008maps}. Since then, drawing contour lines (sometimes also called isolines) has been a frequently used method for displaying 3D surfaces in print. In cartography, the number of contours, or the distance between them, is often chosen depending on the intended purpose of the map. The more contours that are drawn the greater detail can be extracted from the map, but the more information is also needed when drawing the map. Thus, the uncertainty in the information about the topography limits the number of contours that can be drawn. When contour maps are used to display estimated surfaces in modern spatial statistics, the number of contours used are meant to reflect the uncertainty in the estimate. Intuitively, one should be allowed to draw many contours if the uncertainty of the estimated surface is low and fewer contours if the uncertainty is high. However, it has not been clear how the number of contours should be decided nor at what levels the contours should be drawn in order to accurately reflect the variability in the estimated surface.

Even though contour maps are widely used, relatively little research has focused on quantifying their statistical properties. This problem is clearly related to the problem of finding uncertainty regions for contour curves, which was studied by \cite{Lindgren95} using theory of level crossings in Gaussian processes and later by \cite{wameling2003} using conditional simulations. Both these methods construct the uncertainty regions locally, without any control over the simultaneous coverage probabilities for the regions. This problem was recently addressed by \cite{french2014}, using a simulation-based approach, and by \cite{bolin12} using a unified approach for both the contour uncertainty problem and the related problem of finding excursion sets of random fields. To our knowledge, the problem of deciding how many contour curves one should draw in a contour map has previously only been studied by \cite{Polfeldt99}, who proposed a method which we will come back to in Section~\ref{sec:partial}.

The main contributions of this work are threefold. Firstly, we discuss different ways in which the statistical properties of contour maps can be understood. Based on these interpretations, we propose statistical quality measures that can be used to evaluate how appropriate a given contour map is for a random field. Secondly, we propose a strategy for deciding the number of contour levels to use in a contour map, and discuss how the actual contour levels should be chosen. Finally, we discuss the problem of discrete and continuous domain interpretations for contour maps and propose practical methods for interpolating discrete-domain calculations to continuous domains. This is an important topic, because the domain of interest for the analysis often is a continuous subset of $\R^2$, such as the unit square. The statistical computations have to be carried out for a finite number of locations, such as those on a regular lattice covering the domain. Thus, the contour map has to be constructed using some form of interpolation of discrete-domain calculations. The problem of analyzing a single contour curve for a surface obtained by linear interpolation of a kriging prediction on a lattice was briefly discussed by \cite{wamerling02gis} and the interpolation problem has to our knowledge not been studied at all for general contour maps before.

Most of the theory that is presented later is applicable to any random field model. However, we need to restrict ourselves if we also want a computationally efficient method for applying the theory. To that end, we will focus on the class of latent Gaussian models, which is a popular model class that includes many of the standard models in spatial statistics \citep{rue09}. A latent Gaussian model is specified hierarchically using a possibly non-Gaussian likelihood for the data $\mv{y}$ conditionally on a latent Gaussian field $x(\cdot)$. The model parameters, such as correlation ranges and measurement noise variances, typically have prior distributions in a Bayesian setting, but whether or not priors are used does not affect the methods we will discuss here. The quantity of interest is the distribution of the latent field $x(\cdot)$ given the data, which typically is summarized using the posterior mean $\pE[x(\mv{s})|\mv{y}]$ as a point estimate (often called the kriging predictor in geostatistics) and the posterior variances $\pV[x(\mv{s})|\mv{y}]$ as a measure of uncertainty, or using a contour map of the posterior mean.

The structure of the paper is as follows. In Section~\ref{sec:problem}, we introduce some needed notation and give a precise definition of contour maps. The section then contains three subsections that contain the main results of the paper: Section~\ref{sec:partial} introduces the method by \cite{Polfeldt99} and generalizes it to a a new method for quantifying and displaying the uncertainty in a contour map. Section~\ref{sec:measures} introduces three different quality measures for contour maps, based on three different statistical interpretations of contour maps. Finally, Section~\ref{sec:optimal} discusses how the contour levels can be chosen. Section~\ref{sec:practical} presents the practical aspects of how to compute the quality measures for discrete domains and presents a method for interpolating the discrete domain calculations to a continuous domain. Section~\ref{sec:simulations} demonstrates the methods using simulated data and Section~\ref{sec:applications} shows an application to temperature
estimation. Finally, Section~\ref{sec:conclusions} concludes with a
discussion. All methods presented in this work are available in the
\textsf{R} \citep{Rteam13} package \texttt{excursions}
\citep{excursions}. The outline of the US was obtained with the \texttt{maps} package \citep{maps} and the Bayesian inference was performed using the \texttt{INLA} package \citep{lindgren2015software,lindgren10,rue09,martins2013bayesian}.

\section{Contour maps for random fields}\label{sec:problem}
In order to quantify uncertainty of contour maps for random fields,
we first need to more clearly define what such a contour map means.
Only then can questions about formal uncertainty quantification and
appropriate numbers of contour levels be considered.  One of the most
important points is that in presenting results from spatial statistical
inference, contour maps are based on spatial point estimates $f(\cdot)$, or functionals,
of properties of the random field distribution, such as the posterior
mean $\pE[x(\cdot)|\mv{y}]$, and not on realizations of the field itself.  The contour map
is thus also itself a functional of the distribution of the
field.  There is no single choice of functional $f(\cdot)$ to base the
contour map on, and the choice affects the interpretation of both the
contour map itself, and any associated uncertainty measure for it.

Let $\Omega$ be the domain of interest for the analysis. The results
will depend on joint properties of the stochastic process across the
entire domain, so the choice of $\Omega$ is important for the
interpretation of the results. In order to avoid unnecessary
theoretical complications, we assume that $\Omega$ is an open set with
a well-defined area $|\Omega|<\infty$. If the domain of interest is
not open, we take the interior and ignore the behavior on the
boundary.

First, we revisit some notation from \cite{bolin12} for excursion sets and contour sets of fixed functions $f(\s)$, $\s\in\Omega$.  This notation is also applicable to the true contours of realizations of the random field $x(\s)$, $\s\in\Omega$.
\begin{defn}[Excursion and contour sets for functions]
Given a function $f:\Omega\mapsto\R$, the positive excursion set
for a level $u$ is defined by $A_u^+(f) = \{ {\s}\in\Omega; f({\s})>u \}$. The negative excursion set is similarly defined by
$A_u^-(f) = \{ {\s}\in\Omega; f({\s})<u \}$. The contour curve $A_u^c$ for a level $u$ is given by $A_u^c(f) = \left(A_u^+(f)^o\cup A_u^-(f)^o\right)^c$,
where $B^o$ denotes the interior of the set $B$ and $B^c$ denotes the complement.
\end{defn}

It is important to note that we do not define the contour curve (which
is a set in general) $A_u^c$ as the set of locations $\{\s: f(\s) =
u\}$, but rather as the set of all continuous and discontinuous level
$u$ crossings. This is different from how, for example
\cite{french2014} defines contour curves, and the reason for using
this slightly more complicated definition is that we want the theory
to be applicable also for discontinuous fields. Such fields occur
frequently in environmental statistics, for example as soon as we have
a discontinuous covariate for the mean value of the field. The contour
set definition is applicable even to realizations of exotic random
fields such as completely independent Gaussian noise processes, where
the contour set for any finite level $u$ becomes the entire domain,
$\Omega$.  Contours for such random fields are not very useful, so
without any practical limitation we can restrict the analysis to
functions and realizations of random fields that are piecewise
continuous, i.e.\ the probability that the set of discontinuity points
is a null set is $1$.  Also noteworthy, but producing no practical
problems, is that in the presence of discontinuities, contour sets for
different levels are not necessarily distinct.

We can now give a precise definition
for contour maps of a function $f(\cdot)$.
\begin{defn}[Contour maps]\label{def:contourmaps}
Let $f:\Omega\mapsto\R$ be a function. A contour map with $K$ contour levels $u_1 < u_2 < \ldots < u_K$ is defined as the collection of contour curves $A_{u_1}^c(f), \ldots, A_{u_K}^c(f)$ and associated level sets $\mapset_k = \{\s : u_{k}< f(s) < u_{k+1}\}$, for $0\leq k\leq K$, where we define $u_0 = -\infty$ and $u_{K+1}=\infty$. We denote this contour map by $C_f(u_1,\ldots, u_K)$.
\end{defn}

\begin{figure}[t]
\begin{center}
\includegraphics[height=0.25\linewidth]{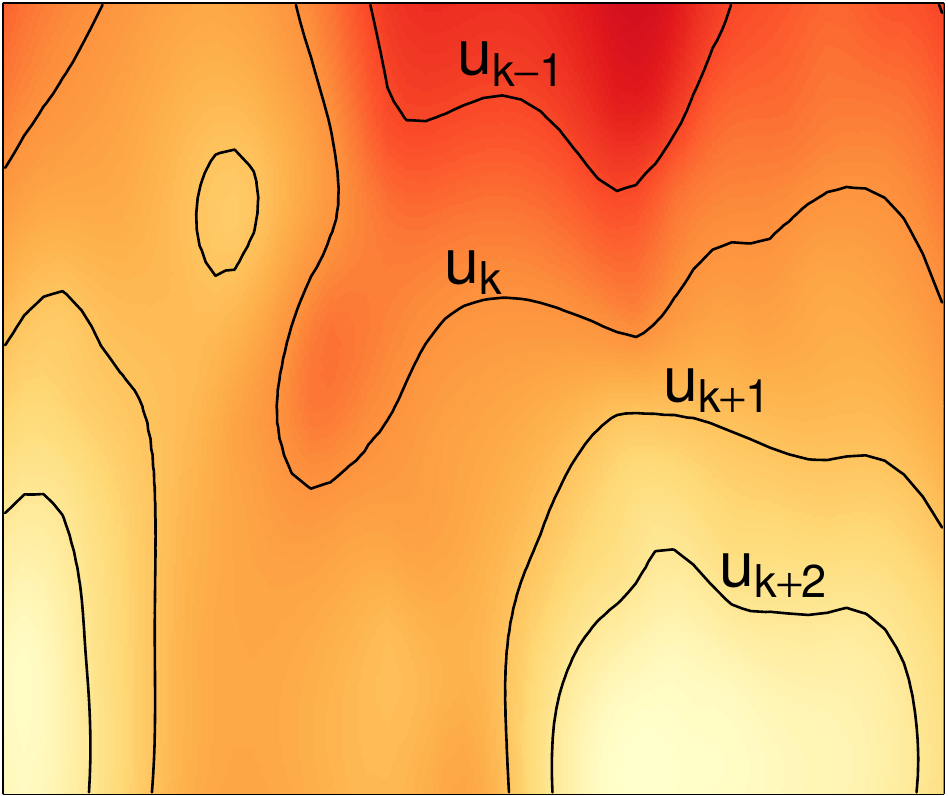}%
\hspace*{2mm}%
\includegraphics[height=0.25\linewidth]{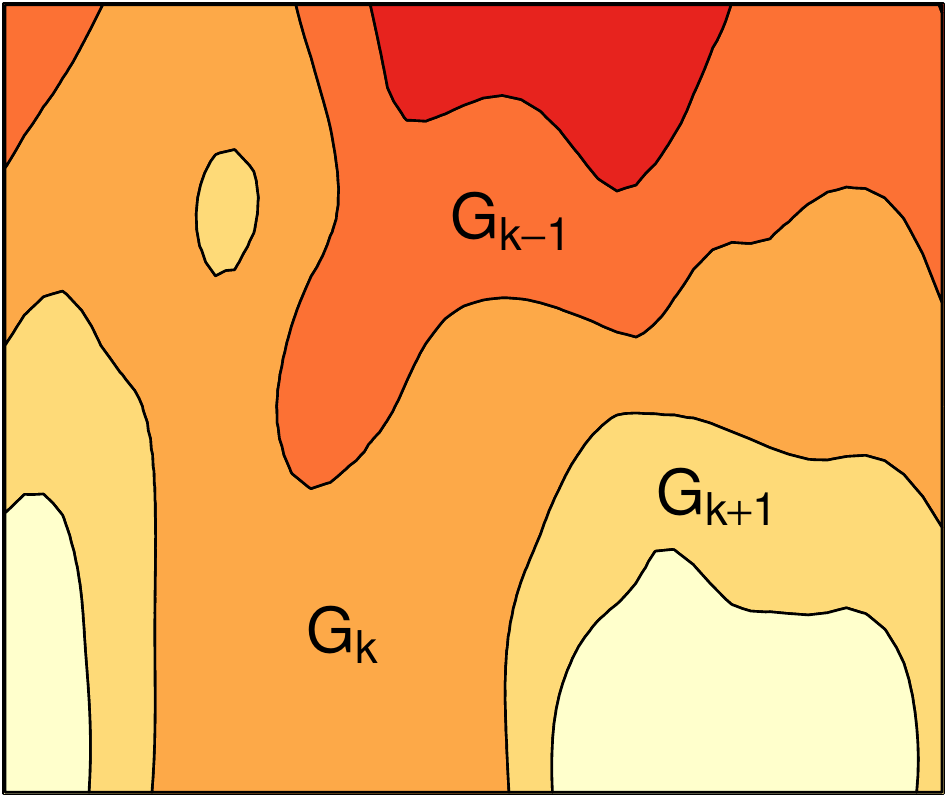}%
\hspace*{2mm}%
\includegraphics[height=0.25\linewidth]{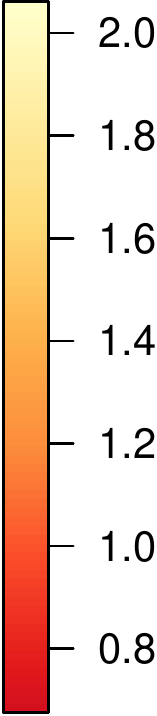}%
\end{center}
\vspace{-0.7cm}
\caption{A function $f(\cdot)$ and its contour curves for levels $u_{k-1}$, $u_k$, $u_{k+1}$ and $u_{k+2}$ (left), and the associated contour map $C_f(u_1,\dots,u_K)$ (right), with the level sets $G_{k-1}$, $G_k$, and $G_{k+1}$ labeled. In the right panel, the color for each region $G_k$ is chosen as the color corresponding to the level $u_k^e = (u_{k+1}-u_k)/2$.}
\label{fig:illustration}
\end{figure}
An illustration of contour curves and the associated contour map for a
specific function $f(\cdot)$ is given in
Figure~\ref{fig:illustration}, showing the level sets $G_k$.
Further details on how the different definitions affect the
interpretation of the contour sets and maps on continuous domains are
given in Section~\ref{sec:continuous}.

We now turn to the problem of interpreting contour maps of functions $f(\cdot)$ as information about a random field $x(\cdot)$.  Most commonly, a contour map for a random field $x(\cdot)$ is constructed by taking $f(\s)$ to be the posterior mean $\pE[x(\s)|y]$ of the field, but other contour maps can be obtained from the posterior median, or some other fixed quantity. The generated contour map does not by itself contain information about the variability in $x(\s)|y$. Therefore, we will now focus on ways to measure how appropriate a given contour map is for a given random field.
Even though we primarily have the situation of looking at a contour map for a posterior distribution in mind, we will simplify the notation by dropping the dependence on the data and consider the more general problem of quantifying how appropriate a contour map $C_{f}(u_1,\dots, u_K)$ is for some general (typically non-stationary) random field $x(\cdot)$.

\subsection{Probabilistic contour map functions}\label{sec:partial}
An intuitively natural interpretation of a contour map for a random field $x(\cdot)$ is that $x(\s)$ should stay between $u_k$ and $u_{k+1}$ for all $\s \in \mapset_k$, for every $k=0,\dots,K$. With the interpretation, the joint probability $\pP(u_{k} < x(\s) < u_{k+1}, \s\in \mapset_{k}, 1\leq k\leq K)$ should be large if the contour map is appropriate for the random field. It is therefore tempting to take this joint probability as a quality measure. Unfortunately, this probability will be low, or even zero in most cases, since the marginal probabilities $\pP(u_k < x(\s) < u_{k+1})$ for any $\s$ close to the boundary of the set $\mapset_k$ will be small. Hence, this interpretation is too strong in general and we need different interpretations of the contour map in order to construct meaningful quality measures.

In order to seek a practical method for determining an appropriate contour level spacing, \cite{Polfeldt99} proposed a method based on analyzing the marginal probabilities
\begin{equation}\label{eq:pmarginal}
p(\s) = \pP(u_{k} < x(\s) < u_{k+1}), \mbox{ for $k$ s.t.~$\s\in \mapset_k$},
\end{equation}
for a Gaussian random field with mean value function $\mu(\mv{s})$ and standard deviations $\sigma(\mv{s})$. In our notation, Polfeldt's analysis was based on the fact that if the distribution for $x$ is Gaussian, these probabilities can be written as
\begin{equation}\label{eq:ps}
p(\s) = \Phi\left(\frac{(1-r(\s))(u_{k+1}-u_k)}{\sigma(\s)}\right) - \Phi\left(-\frac{r(\s)(u_{k+1}-u_k)}{\sigma(\s)}\right),
\end{equation}
where $\Phi$ denotes the standard Gaussian distribution function and
\begin{equation*}
r(\s) = \frac{\mu(\s)-u_k}{u_{k+1}-u_k}.
\end{equation*}

The ratio $(u_{k+1}-u_k)/\sigma(\s)$ is generally dependent on the spatial location $\s$. However, one can compute a common ratio $q = (u_{k+1}-u_k)/\sigma_0$ for the entire contour map if $\sigma(\s)$ can be approximated by a constant $\sigma_0$ and if the contour levels $u_k$ are equally spaced (in the sense that $u_{k+1}-u_k$ is independent of $k$). If $\sigma(\s)$ is replaced by $\sigma_0$ also in \eqref{eq:ps}, one can plot $p$ as a function of $r$ for different values of $q$. Because $r$ measures how close one is to a level contour, $p(\s)$ will always be close to $0.5$ for $r$ close to zero, but it will quickly increase towards 1 if $q$ is chosen high enough.

Polfeldt argued that $x(\s)$ falls within the contour bands for most values of $\s$ if $p(r)$ is close to one for most values of $r$, and one should therefore choose the number of contours so that $q$ is large enough for this to occur. Because $q$ is determined by the spacing between the contours and the average standard deviation of the kriging estimator, this is a reasonable conclusion.

However, there are many approximations in Polfeldt's argument: the strategy only works for Gaussian posterior distributions (although it is straightforward to extend it to other distributions), it assumes that the kriging errors are well-approximated by a constant, and it does not take the spatial dependency of the data into account. In practical applications, these approximations are often difficult to justify, and we will therefore now focus on extending this method to relax (actually remove) the assumptions. The idea we will use is to extend the excursion functions introduced by \cite{bolin12} to contour maps.

In order to extend the method, we first need some notation. The following definition of the contour-avoiding set for a random field is taken from \cite{bolin12}.

\begin{defn}[Contour-avoiding set]\label{def:avoiding}
Let $x({\s})$, ${\s}\in\Omega$ be a random field, let $1-\alpha$ be a predefined error probability, and let $u$ be the value for a contour curve. The $1-\alpha$ contour-avoiding set is defined as the union $\exset{u,\alpha}{}(x) = M_{u,\alpha}^+(x) \cup M_{u,\alpha}^-(x)$. Here,
  \begin{equation*}
(M_{u,\alpha}^+(x), M_{u,\alpha}^-(x)) = \argmax_{(D^+,D^-)}\{|D^-\cup D^+| :
\pP(D^- \subseteq A_u^-(x),\, D^+ \subseteq A_u^+(x))
\geq 1-\alpha\},
  \end{equation*}
is the pair of joint contour $u$ excursion sets, where the sets $(D^+,D^-)$ are open and $|D|$ denotes the area of the set $D$.
\end{defn}

Recall that $D^- \subseteq A_u^-(x)$ means that $x(\s)<u$ for $\s\in D^-$, and that $D^+ \subseteq A_u^+(x)$ means that $x(\s)>u$ for $\s\in D^+$. Thus, the pair of contour-avoiding sets is the largest pair of sets where, with probability $1-\alpha$, one has $x(\s)>u$ in one set and $x(\s)<u$ in the other.

A credible region for the level $u$ contour curve is defined as $\exset{u,\alpha}{c}(x) = (\exset{u,\alpha}{}(x))^c$, which is the smallest set such that with probability at least $1-\alpha$, all level $u$ crossings of $x$ are in the set. Thus, this credible region has a well-defined global interpretation, and one should note that the definition is different from some other definitions for contour uncertainty regions in the literature, e.g.~that of \cite{Lindgren95}. Now, because we are interested in joint properties of contour maps, we extend this definition to multiple contours as follows.
\begin{defn}
Let $x(\s)$, $\s\in\Omega$ be a random field, let ${\uu} = (u_1,\ldots,u_K)$ where $u_1<u_2<\ldots<u_K$ are some predefined levels, and let $1-\alpha$ be a given probability. The collection of joint contour-avoiding sets for the levels ${\uu}$,
${M}_{{\uu},\alpha} = (M_{\uu,\alpha}^1,\ldots,M_{\uu,\alpha}^K)$, is then given by
\begin{equation*}
{M}_{{\uu},\alpha} = \argmax_{(D_1,\ldots,D_K)}\left\{\sum_{k=0}^K |D_k| : \pP\left(\bigcap_k\{D_k \subseteq A_k(x)\}\right)\geq 1-\alpha\right\},
\end{equation*}
where the sets $D_k$ are disjoint and open and $A_k = A_{u_{k+1}}^-(X)\cap A_{u_k}^+(X) = \{s: u_{k} < x(\s) < u_{k+1}\}$. The joint ${\uu}$ contour-avoiding set is given by the union of these sets,
\begin{equation*}
C_{{\uu},\alpha}(x) = \bigcup_k M_{\uu,\alpha}^k
\end{equation*}
\end{defn}

Recall that $D_k \subseteq \mapset_k(x)$ means that $u_k<x(\s)<u_{k+1}$ for $\s\in D_k$. Thus, the set $C_{{\uu},\alpha}(x)$ is the largest set so that with probability at least $1-\alpha$, the random field $x$ satisfies the natural interpretation of the contour map for the locations in the set. Hence, we would like to choose the number of contour lines so that $C_{{\uu},\alpha}(x)$ is large. This procedure is therefore a natural extension of the Polfeldt procedure.

Now, we can go further and calculate a variant of the excursion functions introduced by \cite{bolin12}. The contour map function,
\begin{equation*}
F_{{\uu}}(\s) = \sup\{1-\alpha; \s\in C_{{\uu},\alpha}\},
\end{equation*}
is obtained for each $\s$ by identifying the smallest $\alpha$ such that $\s$ is contained in the contour-avoiding set $C_{\uu,\alpha}$.  The function $F_\uu(\s)$ takes values between zero and one and can be used to visualize the $\alpha$-indexed family of functions $C_{{\uu},\alpha}$. Note that each $C_{{\uu},\alpha}$ can be retrieved as the $1-\alpha$ excursion set of the function $F_{{\uu}}(\s)$. Visualizing this function is again a natural extension of Polfeldt's procedure of visualizing $p(\s)$ as a function of $r$, because we want to choose the number of contours such that $F_{{\uu}}(\s)$ increases rapidly when we move away from the contour lines.

\subsection{Global quality measures}\label{sec:measures}
We will call a function $P(x,C_f)$ a contour map quality measure if it takes values in $[0,1]$ and if $P\approx 1$ indicates that $C_f$, in some sense, is appropriate as a description of the distribution of the random field $x$, whereas $P\approx 0$ indicates that $C_f$ is inappropriate.  Given the contour map function, we define a simple contour map quality measure, $P_0$, as the normalized integral of the contour map function,
\begin{equation}\label{eq:P0measure}
P_0(x,C_f) = \frac1{|\Omega|}\int_{\Omega} F_{{\uu}}(\s) \md \s.
\end{equation}
Loosely speaking, this measure tells us the percentage of the total area for which the statement of the contour map holds.

By revisiting which sets are involved in the joint probability calculations, we will construct two additional quality measures, with geometrically interpretable properties.

\subsubsection{The $P_1$ quality measure}
One possibility is to require that the unions $\mapset_{k-1} \cup \mapset_{k}$ with high probability should contain all level $u_k$ crossings of the process $x$. We define the contour map measure $P_1$ as the probability for this occurring.
To ensure that all values of $x(\cdot)$ are covered by crossing levels, and for ease of notation, we define $u_{j} = -\infty$ for $j\leq 0$ and $u_j = \infty$ for $j>K$. Because $\cup_{l=0}^{k-1} \mapset_l = A_{u_k}^-(f)$ and $\cup_{l=k}^{K} \mapset_l = A_{u_k}^+(f)$, we can write the measure as
\begin{equation*}
P_1(x,C_{f}) = \pP\left(\bigcap_{k=0}^{K}\{x(\s) < u_{k}, \s\in A_{u_{k-1}}^-(f)\}\cap \{x(\s) > u_{k}, \s\in A_{u_{k+1}}^+(f)\}\right).
\end{equation*}
Thus, the interpretation is that if $P_1$ is close to one, then for any $k = 1 ,\ldots, K$ the probability is small for $x$ taking the value $u_k$ outside the two regions $\mapset_{k-1}$ and $\mapset_{k}$ bordering the level $u_k$ contour curve. The following equivalent formulation of the measure is useful for practical calculations,
\begin{equation*}
P_1(x,C_{f}) = \pP\left(\bigcap_{k=0}^{K}\{u_{k-1} < x(\s)< u_{k+2}, s\in \mapset_{k}\}\right).
\end{equation*}
The asymmetry in that the lower bounds are $u_{k-1}$ while the upper bounds are $u_{k+2}$ is caused by the simple fact that each $\mapset_k$ lies between $u_k$ and $u_{k+1}$. More specifically, $\s \in \mapset_{k-1} \cup \mapset_{k}$  is equivalent to that $x(\s) < u_{k}$ for $s \in \mapset_{k-2}$ and that $x(\s) > u_{k}$ for $\s\in \mapset_{k+1}$. This means that  $x(\s) < u_{k+2}$ for $\s \in \mapset_{k}$ and that $x(\s) > u_{k-1}$ for $\s\in \mapset_{k}$, which we can write as $u_{k-1} < x(\s) < u_{k+2}$ for $\s \in \mapset_k$.

From this formulation, we see if $P_1$ is large we have, with high probability, $u_{k-1} < x(\s)< u_{k+2}$ for $\s\in \mapset_k$, i.e.\ that the locations in $G_k$ might even have true values as high as the next level up or as low as the next level down. Thus, this measure is based on a weaker restriction on how the field $x(\s)$ can vary compared to the natural deterministic interpretation that requires the function to lie strictly between two adjacent levels.  For the choice $f(\s)=\pE[x(\s)]$, Figure~\ref{fig:illustration2}(middle) shows five realizations of the level $u_k$ contour curve for $x(\cdot)$, and they lie mostly inside $G_{k-1}\cup G_k$, as desired.

\subsubsection{The $P_2$ quality measure}
\begin{figure}[t]
\begin{center}
\includegraphics[height=0.25\linewidth]{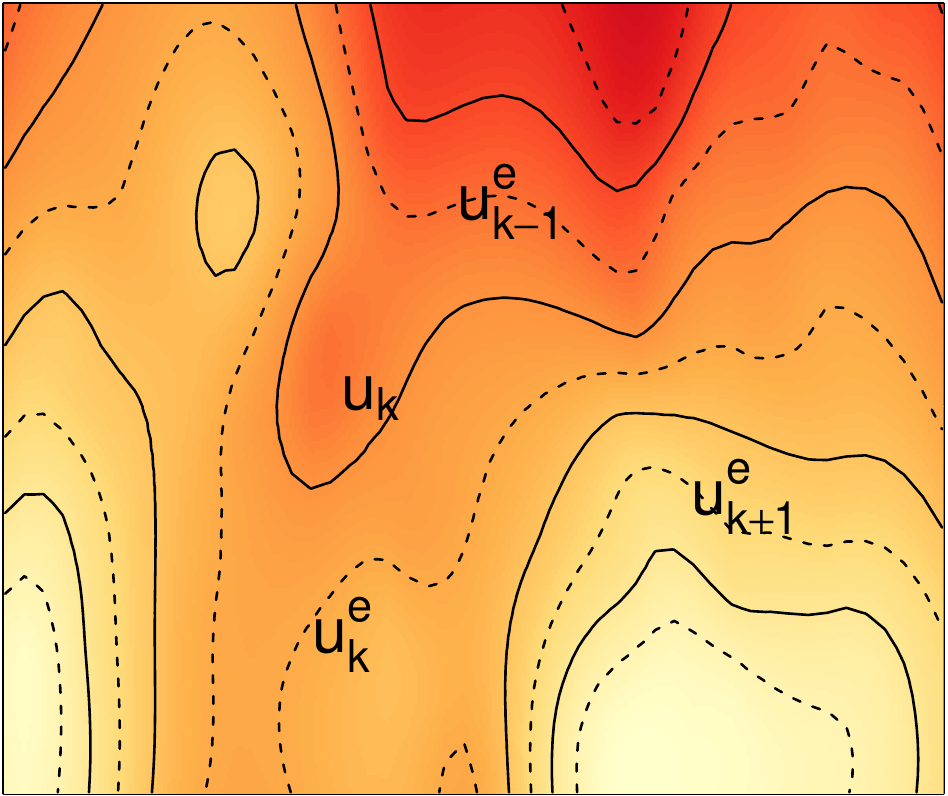}%
\hspace*{2mm}%
\includegraphics[height=0.25\linewidth]{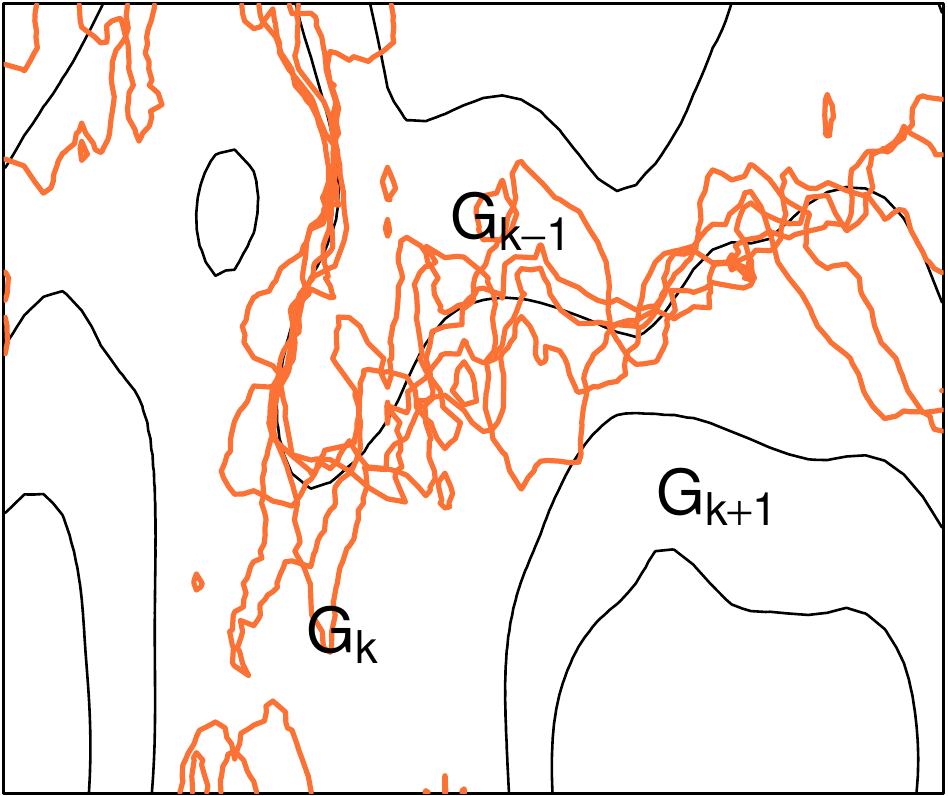}%
\hspace*{2mm}%
\includegraphics[height=0.25\linewidth]{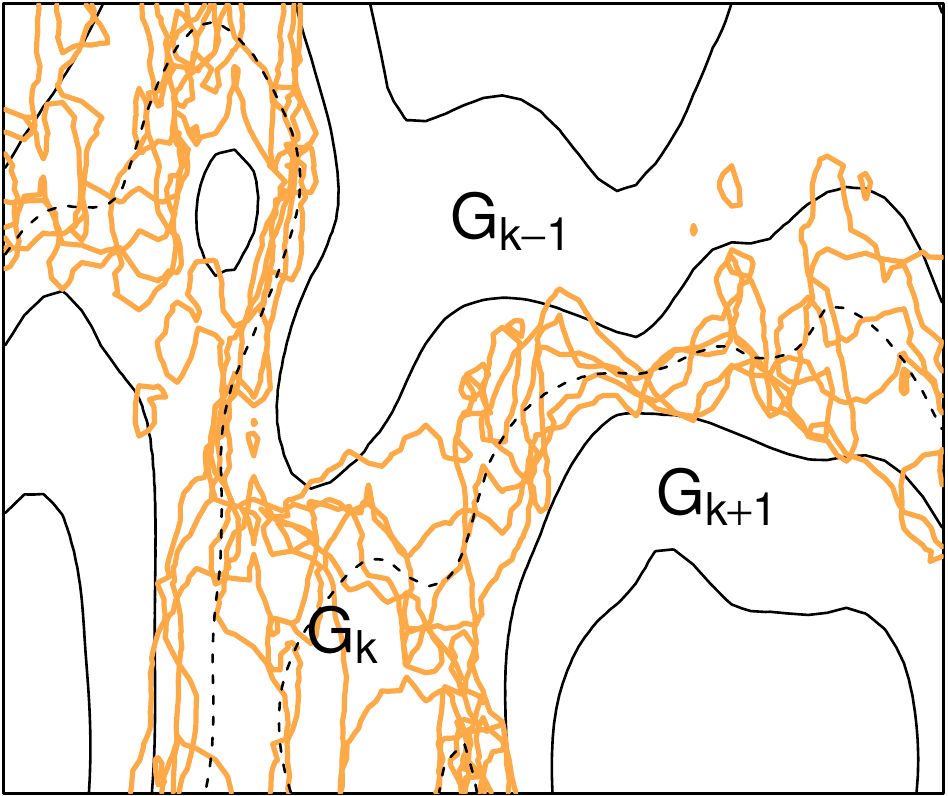}%
\hspace*{2mm}%
\includegraphics[height=0.25\linewidth]{figs/illustration_cbar.pdf}%
\end{center}
\vspace{-0.7cm}
\caption{An expectation function $f(\s)=\pE[x(\s)]$, with contour curves for levels $u_k$, $u_{k-1}^e$, $u_k^e$, and $u_{k+1}^e$ marked (left), five realisations of the $u_k$ contour of $x(\cdot)$ (middle), and five realizations of the $u_k^e$ contour of $x(\cdot)$ (right). In the middle panel, most $u_k$ contours are in the set $G_k \cup G_{k+1}$, with joint probability for all $k$ quantified by the $P_1$ measure. In the right panel, most $u_k^e$ contours are in the set $G_k$, with joint probability for all $k$ quantified by the $P_2$ measure. }
\label{fig:illustration2}
\end{figure}

The $P_1$ focuses in the contour curves, resulting in joint probabilities for overlapping sets $G_{k-1} \cup G_{k}$ for $k = 1, \ldots K$. Another possibility is to focus on the disjoint areas $\mapset_k$ for $k = 0, \ldots, K$. Because $\mapset_k$ contains all locations associated with values between $u_k$ and $u_{k+1}$, it is natural to associate the area $\mapset_k$ with the midpoint level $u_k^e = (u_{k} + u_{k+1})/2$.  To ensure a sensible interpretation for $G_0$ and $G_K$, if $K>1$ we set $u_0 = 2u_1-u_2$ and $u_{K+1}=2u_K-u_{K-1}$ when defining $u_0^e$ and $u_{K}^e$, and if $K=1$ we set $u_0 = u_1 - \sup_{\s\in\Omega} f(\s) + \inf_{\s\in\Omega} f(\s)$ and $u_2 = u_1 + \sup_{\s\in\Omega} f(\s) - \inf_{\s\in\Omega} f(\s)$. The $u_k^e$ values are also used in general when deciding the plotting color for each $G_k$ set, as in Figure~\ref{fig:temp_cm1} and Figure~\ref{fig:illustration}(right). The idea is then that each $\mapset_k$ with high probability should contain all level $u_k^e$ crossings, and we define the contour map measure $P_2$ by the probability for this occurring,
\begin{equation*}
P_2(x,C_f) = \pP\left(\bigcap_{k=0}^{K}\{x(\s) < u_{k}^e, \s\in A_{u_k}^-(f)\}\cap \{x(\s) > u_{k}^e, \s \in A_{u_{k+1}}^+(f)\}\right).
\end{equation*}
This can be interpreted as the simultaneous probability for the level
crossings of $(u_1^e,\,\dots,\,u_K^e)$ all falling within
their respective level sets $(G_1,\,\dots,\,G_K)$, which is then
a simultaneous credible region collection with credibility probability
$P_2(x,C_f)$.  This is close to how one intuitively interprets contour maps.
For the choice $f(\s)=\pE[x(\s)]$, Figure~\ref{fig:illustration2}(right) shows five realizations of the level $u_k^e$ contour curve for $x(\cdot)$, and they lie mostly inside $G_k$, as desired.

If $P_2$ is large, then for any $k = 0, \ldots, K$, the probability is low for $x$ taking the value $u_k^e$ outside the region $\mapset_k$ that is associated with that value. We can also reformulate this measure to an expression more suitable for practical computation,
\begin{equation*}
P_2(x,C_f) = \pP\left(\bigcap_{k=0}^{K}\{u_{k-1}^e < x(\s) <u_{k+1}^e, \s\in \mapset_{k}\}\right),
\end{equation*}
where we interpret $u_{j}^e = -\infty$ for $j < 0$ and $u_j = \infty$
for $j>K$. From this formulation, we see that, with high probability,
$u_{k-1}^e < x(\s)< u_{k+1}^e$ for $\s\in \mapset_k$. Because $u_{k-1} <
u_{k-1}^e$ and $u_{k+1}^e < u_{k+2}$, we have that $P_2$ puts a
stronger restriction on how the field can vary compared with the $P_1$
measure.

All three quality measures $P_0$, $P_1$, and $P_2$ take values between
$0$ and $1$ and can be used to quantify how appropriate a given
contour map is for a given stochastic field.  The $P_0$ measure is
linked to the spatially interpretable contour map function
$F_\uu(\cdot)$, whereas the $P_2$ measure gives the joint credibility
probability for crossings of levels in between the ones displayed in a
standard contour map. The measure $P_1$ is similar in spirit to
$P_2$, but is more permissive.


\subsection{Choosing the contour levels}\label{sec:optimal}
The quality measures presented above can be used to choose an appropriate number of contours. This is done by first deciding on a credibility level, such as $90\%$, and then choosing the largest number of levels, $K$, so that the quality measure for the contour map is above the chosen credibility level.

Given $K$ one must also choose the contour levels $u_1, \ldots, u_K$. When doing this one can either permit arbitrary levels, or mandate a common level spacing, i.e.\ that the difference $u_{k+1}-u_k$ is the same for all $k$.  Given a restriction of using exactly $K$ levels within the range of $f(\cdot)$, one choice, here referred to as Standard, is to place the contour levels
$u_k$ with an even spacing between the minimal and maximal value of
the function $f$.

The Standard choice is the default in \texttt{contour} in \textsc{Matlab}~\citep{Matlab}, whereas the \texttt{contour} function in \textsf{R} \citep{Rteam13} by default uses a more unpredictable heuristic in order to ensure aesthetically pleasing level combinations. More specifically, for a given $K$, the method finds a sequence of about $K+1$ equally spaced values which cover the range of $f$. The values are chosen so that they are 1, 2 or 5 times a power of 10. We refer to this choice as Pretty, and examples of both Pretty maps and Standard maps are shown for simulated data in Section \ref{sec:simulations}.

For Standard contour maps, there is a direct relationship between $K$ and the level spacing. Because the contour levels can cover more than the range of $f$,  this is not the case for Pretty contour maps. Therefore, the level spacing is of more direct interest than $K$ when comparing different contour maps. See Figure \ref{fig:ex2_cm} for examples of pretty contour maps with the same $K$ but with different level spacing.

If the restriction of a common level spacing were to be removed, an alternative would be to choose the non-equally spaced levels that maximize the quality measures subject to the restriction that the levels cover the range of $f$. Computationally, a cheap approximate solution to this can be obtained using the bounds for the $P$ measures presented in Section \ref{sec:discrete}. However, due to the lack of a clear interpretation of the resulting contour maps, we do not pursue this further.

\section{Practical considerations and continuous domains}\label{sec:practical}
We have now presented several quality measures for contour maps, including an extension to the \cite{Polfeldt99} procedure for choosing the number of contour lines. However, these tools would not be of much use unless we also have a method for computing them. Thus, in this section we go through the practical details of how to calculate the different quality measures.

A common scenario for when contour maps are used is when the domain $\Omega$ is a continuous subset of $\R^2$, such as the unit-square. In order to do any calculations in this case, one has to discretize the problem. We therefore start with the simpler problem of when $\Omega$ is discrete in Section~\ref{sec:discrete}. After this, we briefly describe how one can approximate the continuous-domain case using corrections to the discrete-domain computations in Section~\ref{sec:continuous}.

\subsection{Calculating the measures for discrete domains}\label{sec:discrete}
Assume that the process is defined on a discrete domain with $n$ locations $\{\s_1,\dots,\s_n\}$, such as a regular lattice, so that $\mv{x} = \{x_1, \ldots, x_n\}$. Furthermore, let $\pi(\mv{x})$ be the joint probability density function for the process, and let $\pi(x_i)$ denote the marginal density function for location $i$.

Simple upper bounds for the measures $P_1$ and $P_2$ can be obtained using the marginal probabilities for the $x_i$ variables. Let
\begin{align*}
\rho_i^1 &= \pP(u_{j-1} < x_i< u_{j+2}) = \int_{u_{j-1}}^{u_{j+2}}\pi(x_i) dx_i,\\
\rho_i^2 &= \pP(u_{j-1}^e < x_i< u_{j+1}^e) = \int_{u_{j-1}ê}^{u_{j+1}ê}\pi(x_i) dx_i,
\end{align*}
 where $j$ is the index such that $s_i\in \mapset_j$. The joint measures now satisfy $P_1 \leq \min(\rho_i^1)$ and $P_2 \leq \min(\rho_i^2)$. If we are interested in a contour map where $P_k>1-\alpha$, we can use these bounds to quickly reject all contour maps where $\min(\rho_i^k)<1-\alpha$ in order to reduce the number of contour maps we have to compute the actual measure for.

In order to calculate the measures, we need to integrate the joint density of the random field $x$. For example, $P_2$ is calculated as
\begin{equation*}
P_2 = \int_{{a_1}}^{{b_1}}\dots\int_{{a_n}}^{{b_n}}\pi(\mv{x})\md \mv{x}
\end{equation*}
where $a_{i} = u_{j-1}^e$ and $b_{i} = u_{j+1}^e$, and for each $i$, $j$ is the index that fulfills $\s_i\in \mapset_j$. This high dimensional integral is generally difficult to calculate efficiently. However, it can be estimated efficiently using the methods described in \cite{bolin12} for latent Gaussian models with Markov properties, using sequential importance sampling. For latent Gaussian models, the posterior distribution is generally not Gaussian, but the integral can still be estimated efficiently by using, for example, the quantile correction method described in \cite{bolin12}.

In order to compute the $P_0$ measure, we also need to calculate the contour map function $F_\uu(\cdot)$. Because the contour map function is an excursion function, we can again directly use the sequential integration method by \cite{bolin12} when calculating the contour map function. In order to use the sequential method, we need a parametric family for the possible contour avoiding sets for $\uu$. A natural choice of such a parametric family is $D(a) = \{\s: p(\s)>1-a\}$, where $p(\s)$ are the marginal probabilities defined in \eqref{eq:pmarginal}. Thus, for a given value of $a\in[0,1]$, $D(a)$ defines a subset of $\Omega$ which serves as a candidate for the contour-avoiding set. See \cite{bolin12} for further technical details of the method.

\subsection{Discrete and continuous domain interpretations}\label{sec:continuous}

For a problem defined on a discrete spatial graph, the crossings of a
level can be described as the set of graph edges that connect nodes with
values below and above the level.  This does not quite match the usual
concept of a contour curve, which requires the function to be defined
at all spatial locations.  It is therefore important to consider the
link between the discrete computational model and the continuous
spatial domain when interpreting the discrete calculations.

For models representing cell averages on lattices, or other spatial
sub regions, the natural interpretation of the discrete contour map
calculations is to define the contour curves as the cell edges that
separate nodes in different $G_k$ sets.  This is equivalent to
applying Definition~\ref{def:contourmaps} to a piecewise constant
function over continuous space. However, because we want to be able to
handle more smooth functions, we instead consider situations where the
random field part of a model can be discretized with weights for
piecewise linear local basis functions.  Common contour plotting
methods are based on variations of such linear interpolation,
e.g.\ \texttt{contour} in \textsf{R} \citep{Rteam13} and
\textsc{Matlab} \citep{Matlab}. In practical generalized linear
models, these continuous functions are combined with potentially
discontinuous covariates.  The uncertainty about the resulting smooth
level crossings and level-crossing jumps can be handled with more
careful treatment of the sub-cell localization of the contour curves.

\subsubsection{Construction of continuous uncertainty interpretations}
\label{sec:contiuousconstruction}

Given contour map calculations on discrete point locations, the
contour map function $F_\uu(\s)$ for the continuous spatial domain can be
approximated, by assuming smoothness of the random field away from the
contours.  The main idea is to interpolate the discretely computed
$F_\uu(\s)$ probabilities, assuming monotonicity of the random field
between the discrete computation locations.  For models constructed as
piecewise linear basis approximations, this assumption is
automatically fulfilled.

The computational models in Section~\ref{sec:simulations}
and~\ref{sec:applications} use piecewise linear basis representations
of stochastic PDE on triangles~\citep{lindgren10}.
Given the values of $F_\uu(\s)$ on the vertices of a triangulation, the
contribution to the $1-\alpha$ contour for $F_\uu(\s)$ from each triangle $\mathcal{T}=(\s_1,\s_2,\s_3)$ depends on if the triangle corners belong to the same or different $G_k$ level sets.  If they belong to different sets, then the continuous version of $F_\uu(\s)$ must go to zero somewhere inside the triangle, since it must have a contour separating the different $G_k$.  In fact, this could be the case for any triangle, but triangles with common vertex $G_k$ are unlikely to have interior contours. Because the precise location of interior contours are unknown and  difficult to estimate, a practical approach to interpolating $F_\uu(\s)$ while taking $G_k$ into account is to first eliminate all triangles with vertices that do not belong to the same $G_k$ set.  The $1-\alpha$ contour sets for $F_\uu(\s)$ are then found by interpolation within each triangle of the pruned domain. The contour sets are made up of line segments for $1-\alpha$ crossings of $F_\uu(\s)$ on the interior of the triangles, and partial sections of edges on the domain boundary for values of $F_\uu(\s)$ above $1-\alpha$.
Optimal triangle
interpolation based on the joint properties of the entire model is
intractable, but there are several parsimonious options.  For
interpolation to a location $\s=w_1\s_1+w_2\s_2+w_3\s_3$ in the triangle
$\mathcal{T}$, where $\{(w_1,w_2,w_3);\, w_1,w_2,w_3\geq 0,\,
\sum_{k=1}^3 w_k=1\}$ are barycentric coordinates, the options in the
\texttt{excursions} package include
\begin{description}
\item[Step:] $F_\uu(\s) = \min\{F_\uu(\s_1), F_\uu(\s_2), F_\uu(\s_3)\}$,
\item[Linear:] $F_\uu(\s) = \sum_{k=1}^3 w_kF_\uu(\s_k)$, and
\item[Log:] $F_\uu(\s) = \exp\{\sum_{k=1}^3 w_k\log[F_\uu(\s_k)]\}$.
\end{description}
For the Step and Log method, zero-width \emph{needle sets} are avoided
in the set construction by first eliminating any triangles where
$F_\uu(\s)=0$ for any of the vertices.

\begin{figure}
  \centerline{%
  \includegraphics[height=0.12\linewidth,angle=90]{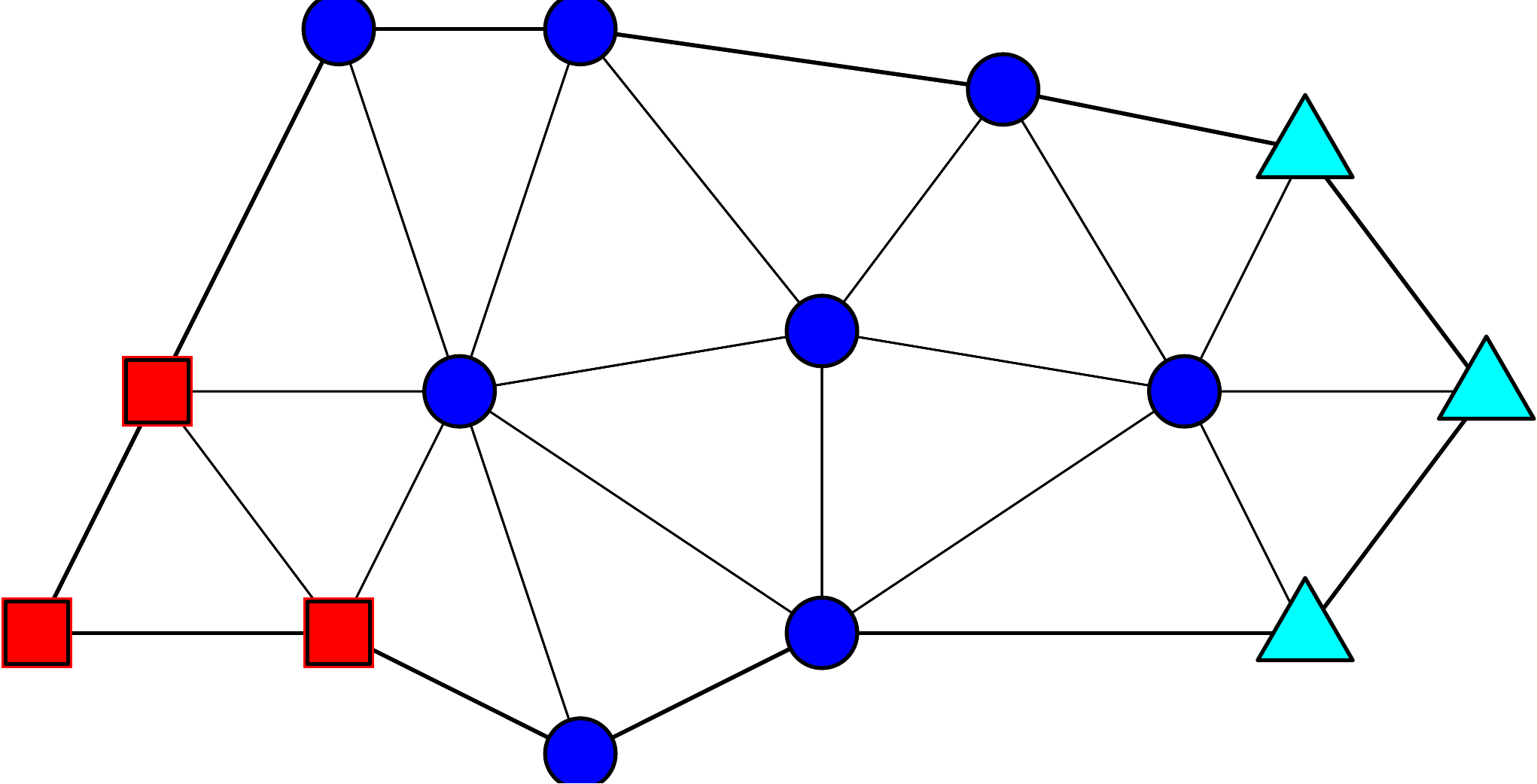}
  \hspace*{2mm}
  \includegraphics[height=0.12\linewidth,angle=90]{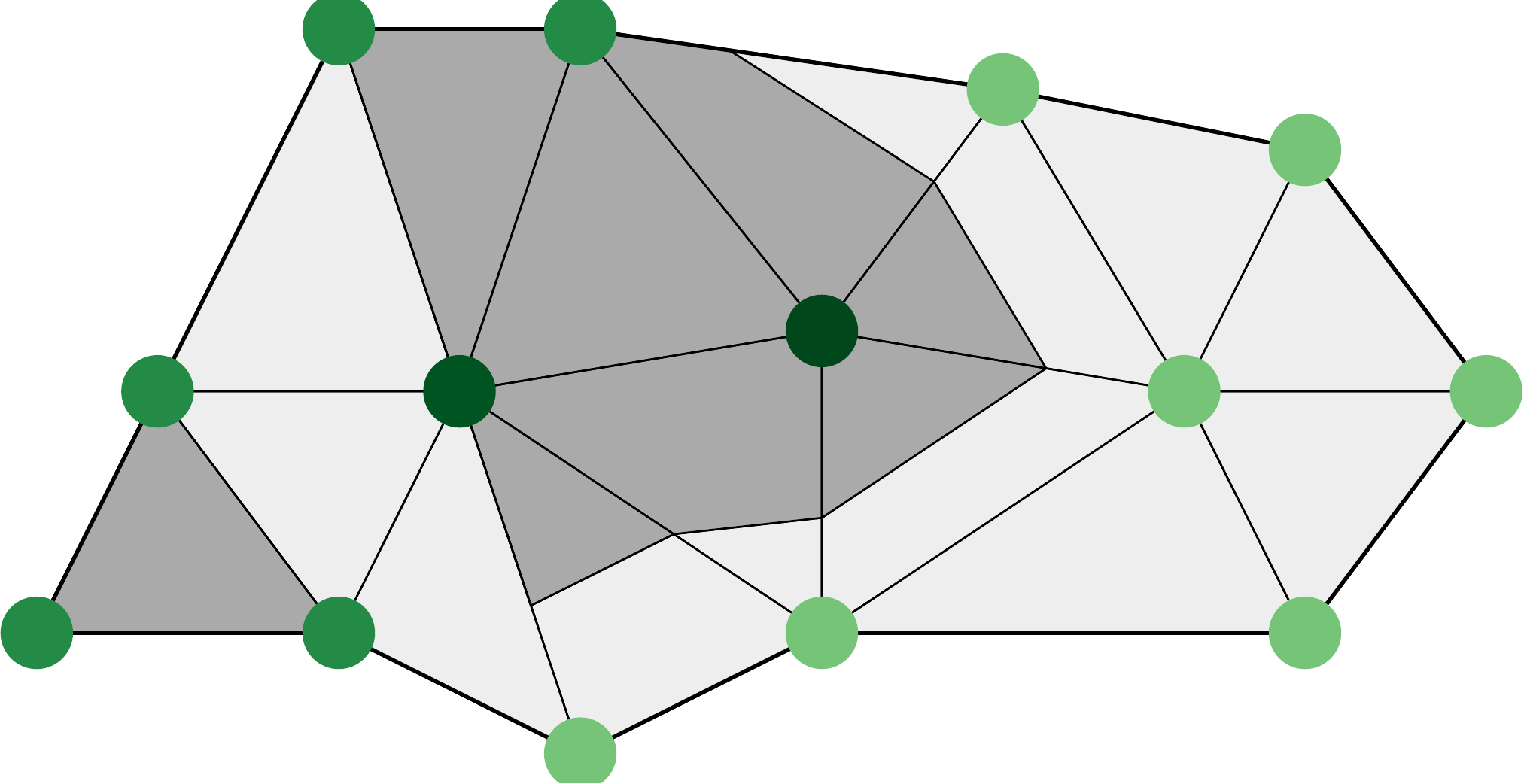}
  \hspace*{2mm}
  \includegraphics[height=0.12\linewidth,angle=90]{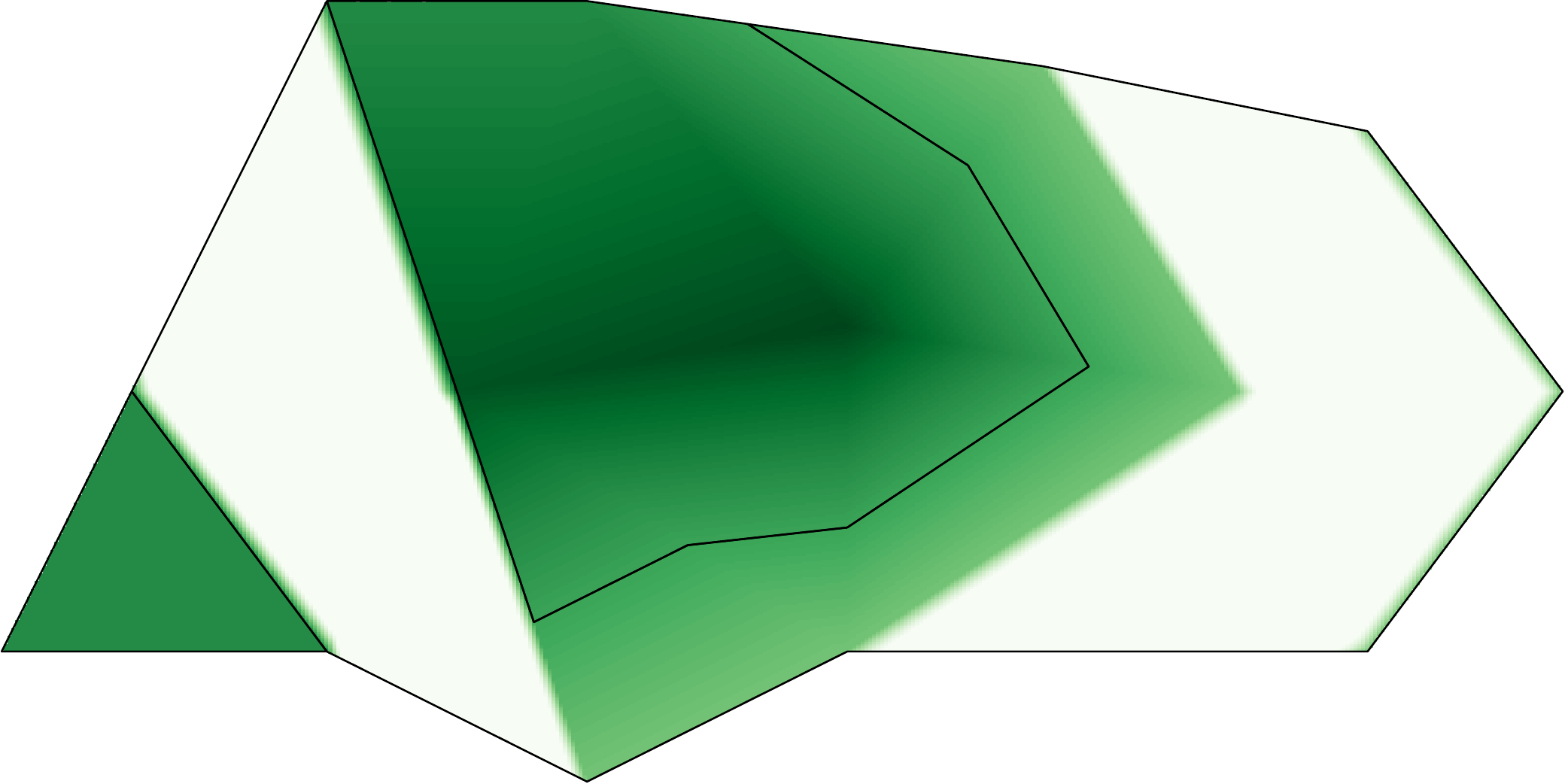}
  \hspace*{2mm}
  \includegraphics[height=0.23\linewidth]{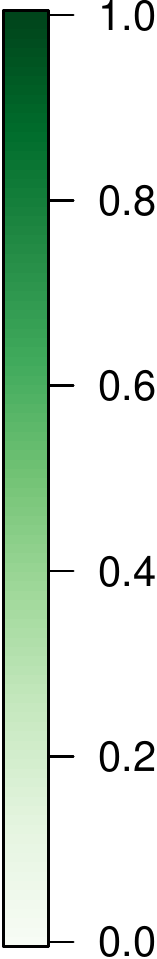}
  }
  \caption{The left panel shows a triangulation with coloured indicators for the levels sets $G_0$ (red squares), $G_1$ (blue circles), $G_2$ (cyan triangles) marked for each vertex. The middle panel shows results of the discrete calculations of $F_{\uu}(\s)$ at the vertices. Interpolation of the values within each triangle according to the algorithm in Section \ref{sec:contiuousconstruction} followed by thresholding yields the contour avoiding sets shown in dark grey. The right panel shows the full interpolated version of $F_{\uu}(\s)$, with the contour-avoiding set outline superimposed.}
  \label{fig:continterp}
\end{figure}

A practical approach for visualizing the interpolated function is to compute a triangle subdivision, splitting each triangle into four new ones in each step, with the chosen interpolation method used to generate the values on the new vertices. Then, linear interpolation on the subdivided triangles, can be used when plotting, as illustrated in Figure~\ref{fig:continterp}.

\begin{figure}[t]
  \centerline{%
    \includegraphics[width=0.32\linewidth]{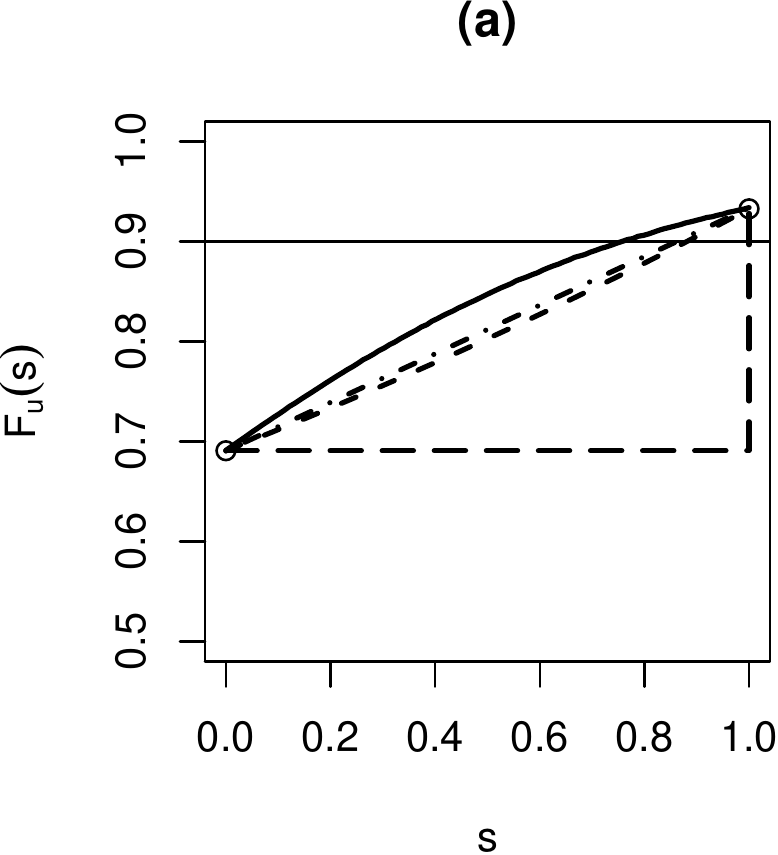}%
    \hspace{0.02\linewidth}%
    \includegraphics[width=0.32\linewidth]{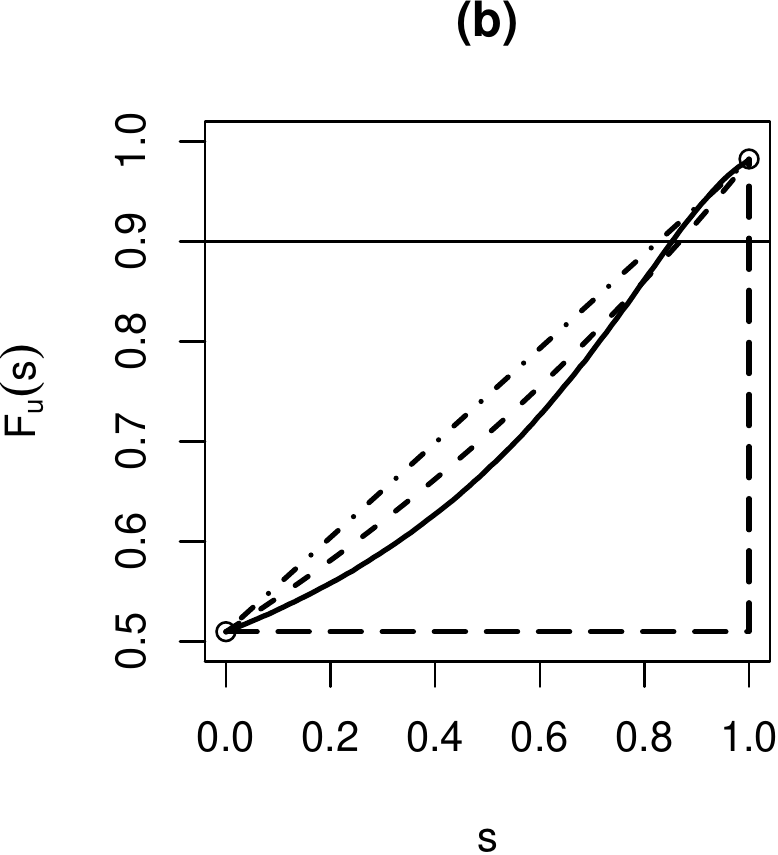}%
    \hspace{0.02\linewidth}%
    \includegraphics[width=0.32\linewidth]{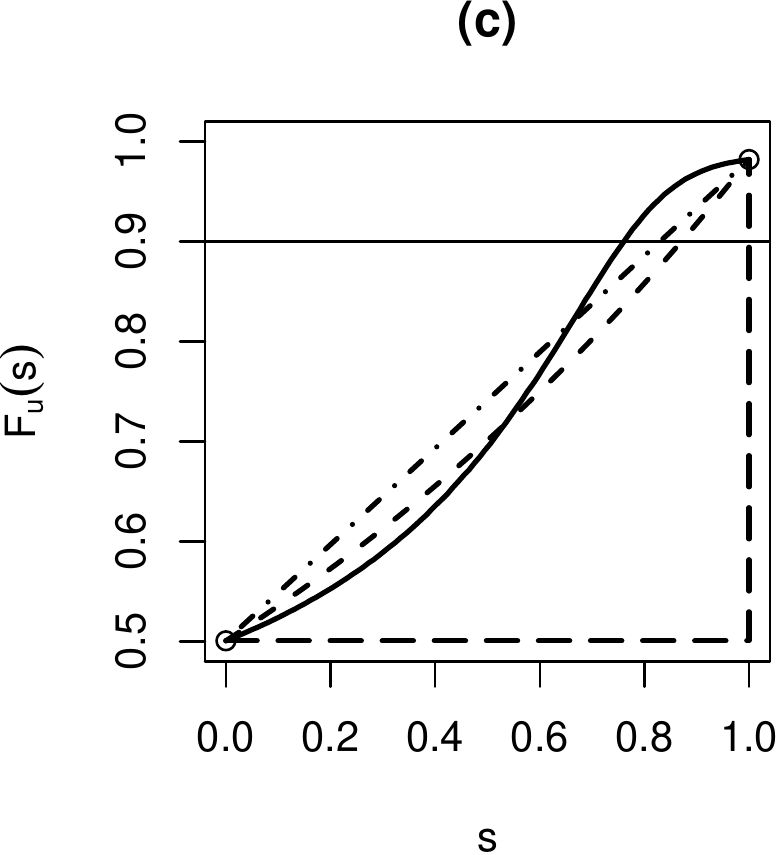}%
  }
  \caption{The true $F_\uu(s)$ (---), and interpolated approximations
    Step (--~--), Linear (--\,$\cdot$), and Log (-\,-), for three test
    cases involving a model discretized at $s=0$ and $s=1$.  A typical target probability $0.9$ is shown for reference. The three test cases (a), (b), and (c) are obtained assuming a Gaussian distribution for $[x(0) \, x(1)]^\top$, where the parameters are varied (see the text in Section \ref{sec:contiuousconstruction} for details) to show typical behaviors of the interpolation methods. One can note that the method Step is always conservative, Log is conservative for large target probabilities, whereas Linear is non-conservative for the target probability in case~(b).}
  \label{fig:interpolation}
\end{figure}

The qualitative behavior of the interpolation methods can be assessed
by considering test cases consisting of only two spatial locations,
which allows exact calculation of $F_\uu(s)$ for a single level $u$
along the line segment between the two locations. We define the joint
distribution for $x(0)$ and $x(1)$ to be
\begin{align*}
  \mat{x(0)\\x(1)}
  &\sim
  \pN\left(\mat{\mu_0\\\mu_1},
  \mat{\sigma_0^2 & \rho\sigma_0\sigma_1 \\
    \rho\sigma_0\sigma_1 & \sigma_1^2}\right) ,
\end{align*}
and define the linear basis representation for $s\in[0,1]$ as
$x(s)=(1-s)x(0)+sx(1)$. The true $F_\uu(s)$ function is compared with
its interpolation approximations for different combinations of
distribution parameters $(\mu_0,\mu_1,\sigma_0,\sigma_1,\rho)$ and
thresholds $u$.  The three test cases shown in
Figure~\ref{fig:interpolation} are
\begin{itemize}
\item[(a)]
  $(\mu_0,\mu_1,\sigma_0,\sigma_1,\rho)=(0,\,1,\,1,\,1,\,0.9)$ with
  $u=-0.5$,
\item[(b)]
  $(\mu_0,\mu_1,\sigma_0,\sigma_1,\rho)=(0,\,2,\,4,\,1,\,0.9)$ with
  $u=-0.1$, and
\item[(c)] $(\mu_0,\mu_1,\sigma_0,\sigma_1,\rho)=(0,\,2,\,4,\,1,\,0)$
  with $u=-0.1$.
\end{itemize}
The interpolation method Step is always conservative, Log is
conservative for large target probabilities, whereas Linear is
non-conservative for the target probability $0.9$ in test case~(b).
When both Linear and Log are non-conservative, Log stays closer to the
true $F_\uu(s)$ function. In this simple example, $x(0)$ and $x(1)$ could be thought of as two of the neighboring locations on a mesh in two dimensions, and no major differences arise if one were to consider 2D interpolation over a triangle in the mesh.

\subsubsection{Numerical assessment of the continuous uncertainty
  interpretations}
\label{sec:numericalassessment}

Coverage probabilities of credible contour regions constructed by
different methods were assessed by \citet{french2014}, for models
where the true fields were on the same spatial resolution as the
discrete models.  The methods from \citet{bolin12} for a single
contour level, which were generalized to multiple levels in
Section~\ref{sec:partial}, were assessed by performing discrete
calculations based on the pointwise values at the lattice cell
centers. Unfortunately, the comparison was designed to assess only the
alternative contour concept from \citet{french2014}, and lead to
severely underestimated coverage probabilities for credible contour
regions based on Definition~\ref{def:contourmaps}.  For an accurate
assessment, one must either evaluate the pointwise calculations with
respect to pointwise level exceedances, or take the properties of a
continuous domain linear basis representation into account, as done in
Section~\ref{sec:contiuousconstruction}.

The numerical tests in Table~\ref{tab:coverage} show that the
interpolation methods are conservative or on target when the
smoothness assumption is fulfilled, but differences appear when the
true random field is defined on a higher spatial resolution, where
segments of the true contours can appear and disappear in between the
discrete calculation points.  More specifically, to assess the
resulting credible contour map coverage probabilities for a single
level $u=0$, we simulated 100 zero mean random fields each from four
different models, with the true model defined at a matching and higher
resolution, and the differentiability $\nu$ of the random field equal
to $1$ and $2$ in the model in Section~\ref{sec:simulations}.  Each
field was observed at $500$ uniformly random locations, with the ratio
between the variances for the field and the observation noise was
fixed to $9$, and this was repeated $10$ times for each true field
realization.  The spatial range was fixed to $3$, with the analysis
domain as a $10\times 10$ square, and the triangulation nodes for the
discretized model were placed in a $20\times 20$ lattice.  The higher
resolution fields were defined on a $200\times 200$ lattice.
\begin{table}[t]
\centering
\begin{center}
\begin{tabular}{lccccc}
\toprule
Resolution & $\nu$ & Step & Linear & Log & Pointwise\\
\cmidrule(r){1-6}
Matching & $1$ &
$0.932(0.026)^{\phantom{+}}$ & $0.895(0.032)^{\phantom{+}}$ & $0.898(0.032)^{\phantom{+}}$ & $0.904(0.031)^{\phantom{+}}$\\
& $2$ &
$0.932(0.026)^{\phantom{+}}$ & $0.887(0.033)^{\phantom{+}}$ & $0.891(0.033)^{\phantom{+}}$ & $0.919(0.029)^{\phantom{+}}$\\
High & $1$ &
$0.850(0.037)$ & $0.583(0.052)^{-}$ & $0.641(0.050)^{-}$ & $0.901(0.031)^{\phantom{+}}$\\
& $2$ &
$0.990(0.010)^{+}$ & $0.959(0.021)^{+}$ & $0.968(0.018)^{+}$ & $0.986(0.012)^{+}$\\
\bottomrule
\end{tabular}
\end{center}
\caption{Estimated coverage probabilities and associated Monte Carlo
  standard deviations for the simulation study in
  Section~\ref{sec:numericalassessment}. The target coverage
  probability was $0.90$, and estimates significantly below or above
  the target are marked with $-$ and $+$, respectively.  For the
  Matching cases, the true field and numerical evaluations were on the
  same spatial scale, and for the High cases, the true field was
  defined on a higher resolution.  The coverage probabilities for
  Step, Linear and Log are with respect to credible regions for
  continuous contour curves, whereas the coverage for the pointwise
  evaluation refers to the contour-avoiding sets of the field
  evaluated at the node locations of the discrete lower resolution
  model.}
\label{tab:coverage}
\end{table}

The resulting Monte Carlo estimates of the coverage probabilities for
the interpolation methods Step, Linear, and Log, are shown in
Table~\ref{tab:coverage}, for a target probability of $0.90$.  For
comparison, the contour-avoiding sets for the nodes of the discretized
model were also evaluated, listed as Pointwise.  When the true field
is defined on a scale matching the discrete calculations, all methods
have coverage probabilities matching the target, regardless of the
smoothness of the continuous limit of the model.  When the true field
is defined on the higher resolution and with $\nu=1$, the Pointwise
evaluations are still on target, but the interpolation methods have
lower coverage, due to not fully taking the within-triangle
variability into account, and Step is the least non-conservative
method, followed by Log, as expected from the test cases in
Figure~\ref{fig:interpolation}.  In contrast, when the true field has
$\nu=2$, all four methods have overinflated coverage probabilities.
This is likely due to the linear basis functions providing a
relatively poorer approximation of the smooth covariance function,
with the variance at each node of the low dimensional representation
being larger than the variance in the continuous limit
\citep[see][]{lindgren10,bolin13comparison}. This gives higher
uncertainty about the locations of crossings, which results in larger
credible regions with higher actual coverage probability than the
target.  To avoid this, the computational resolution needs to be
increased to more closely match the structural scale of the true field
realizations. Further studies are required to give more precise
guidance, but a basic \emph{ad hoc} rule is that the longest edge of
any triangle should be at most one tenth of the spatial correlation
length for models with $\nu=1$ but a quarter to a half may be sufficient for
$\nu=2$. The reason for this is that the true contour curves are rough for models with $\nu=1$ but smooth for models with $\nu=2$.

\section{An example using simulated data}\label{sec:simulations}
In this section, we illustrate the contour map methods using an example with simulated data. Let $x({\s})$, ${\s}\in [0,10] \times [0,10]$, be a mean-zero Gaussian Mat\'ern field. Its covariance function is given by
\begin{equation}\label{eq:matern}
C(\|\mathbf{h}\|) = \frac{2^{1-\nu}\phi^2}{(4\pi)^{\frac{d}{2}}\Gamma(\nu + \frac{d}{2})\kappa^{2\nu}}(\kappa\|\mathbf{h}\|)^{\nu}K_{\nu}(\kappa\|\mathbf{h}\|),
\end{equation}
where $\nu$ is a shape parameter, $\kappa$ a scale parameter, $\phi^2$ a variance parameter, $K_{\nu}$ is a modified Bessel function of the second kind of order $\nu>0$, and $\|\cdot\|$ denotes the Euclidean spatial distance.

\begin{figure}[t]
\begin{center}
\includegraphics[height=0.25\linewidth]{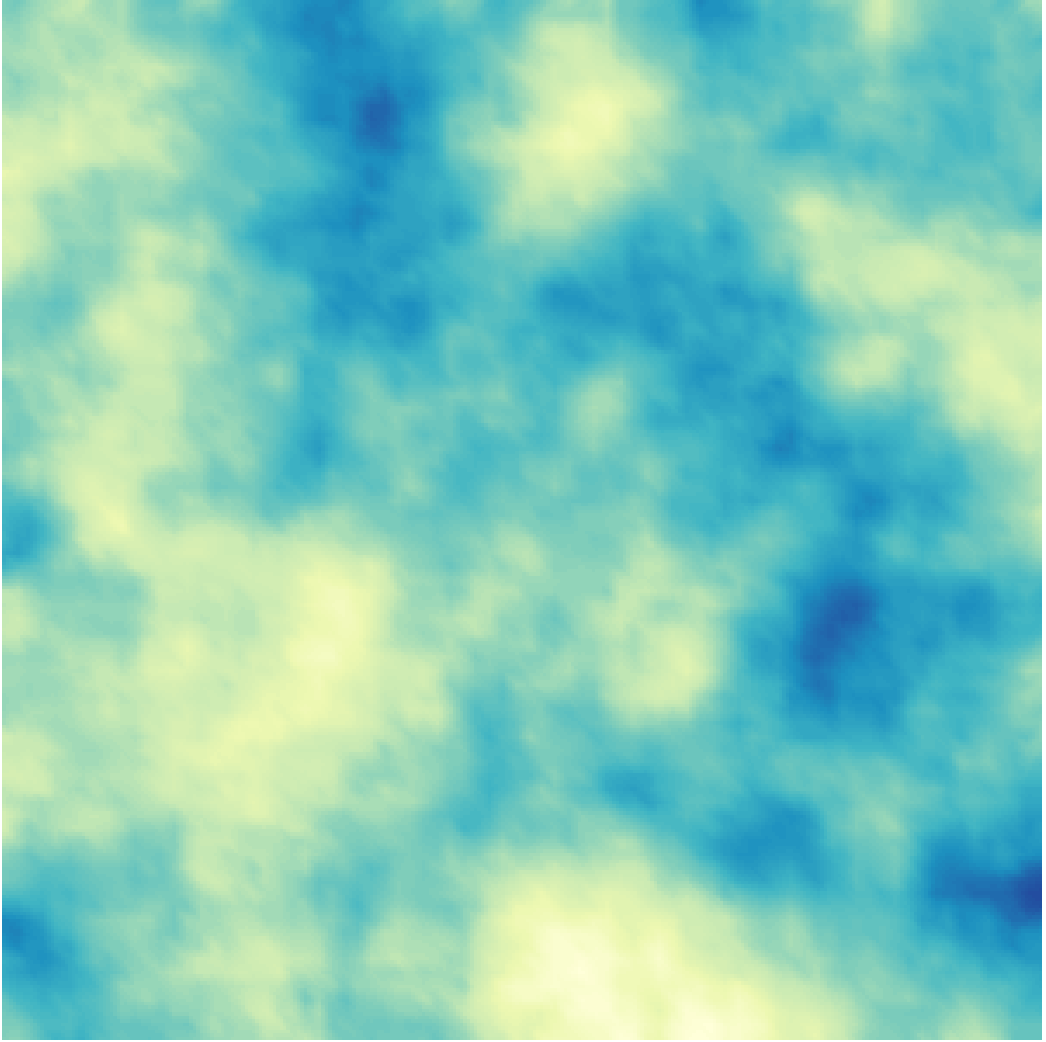}%
\hspace*{2mm}%
\includegraphics[height=0.25\linewidth]{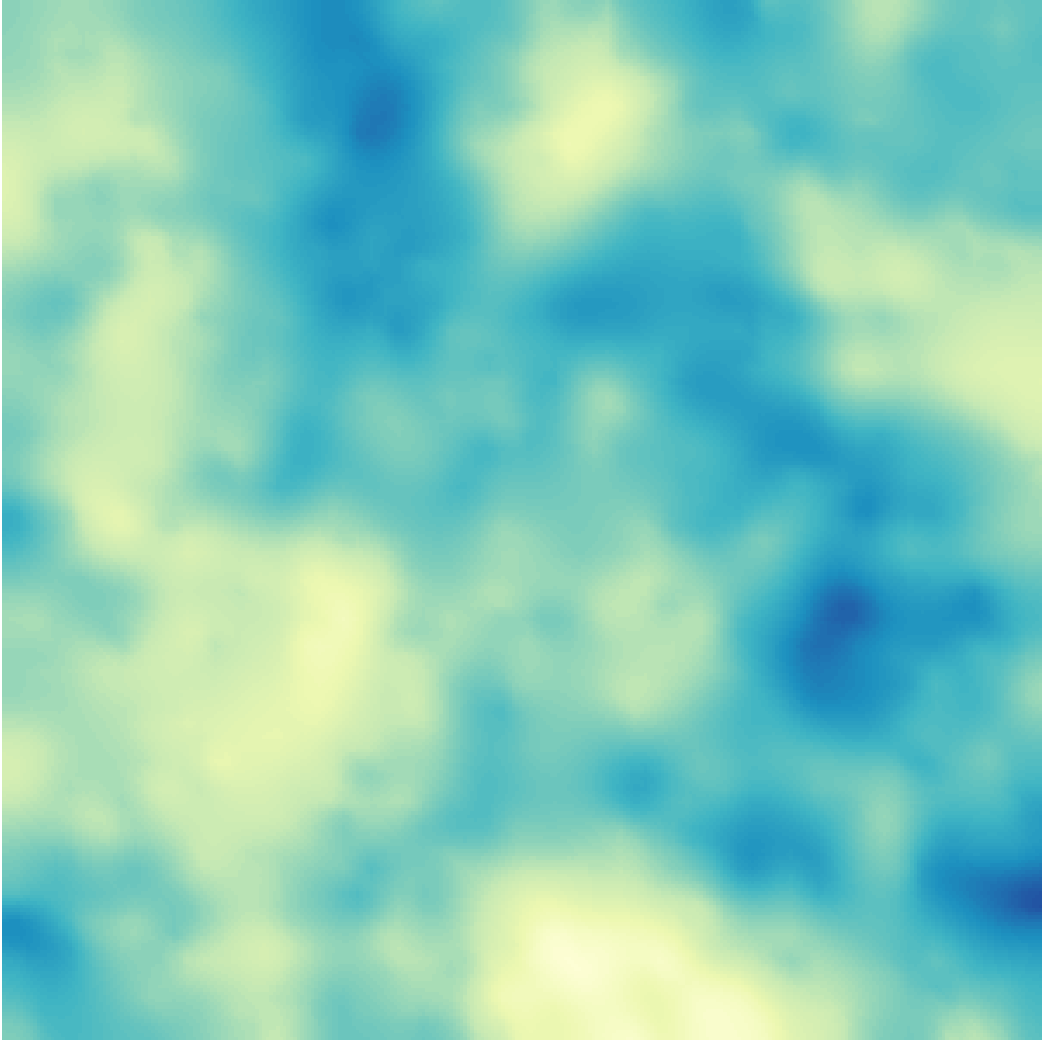}%
\hspace*{2mm}%
\includegraphics[height=0.25\linewidth]{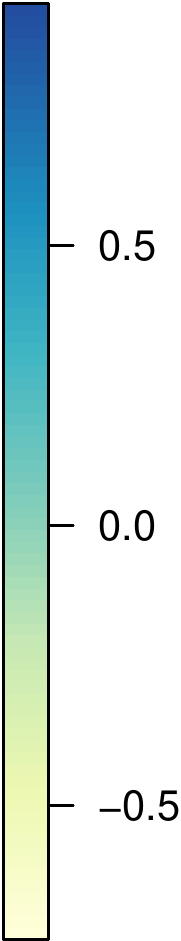}%
\hspace*{2mm}%
\includegraphics[height=0.25\linewidth]{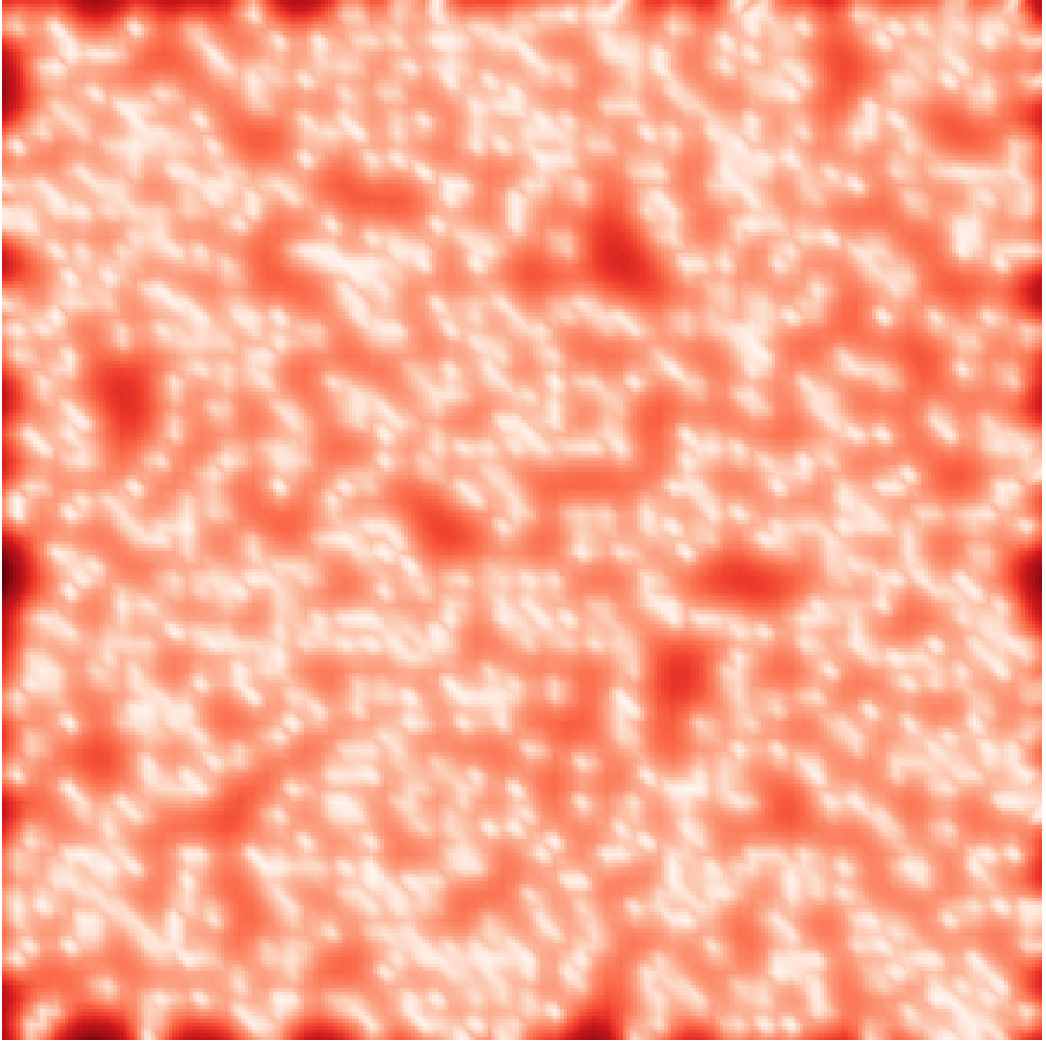}%
\hspace*{2mm}%
\includegraphics[height=0.25\linewidth]{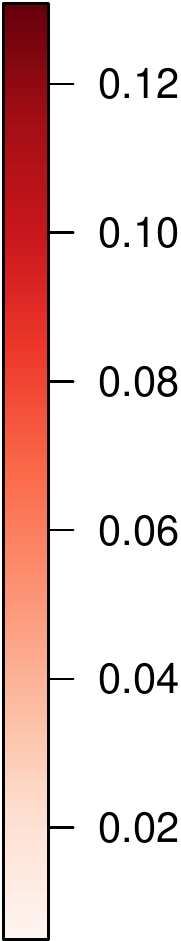}%
\end{center}
\vspace{-0.4cm}
\caption{A simulated Mat\'ern field (left), the kriging prediction of the field based on $500$ measurements (middle), and the corresponding standard errors (right). The uncertainty pattern seen in the standard errors is determined by the spatial pattern of the observation locations, and the kriging point estimate is smoother than the true field. }
\label{fig:ex2}
\end{figure}

We use the SPDE representation by \cite{lindgren10} of the field and generate a realization of the field with parameters $\nu = \phi = \kappa = 1$. The field is observed under additive $\pN(0,0.001^2)$ noise at $1000$ locations chosen at random in the square. Given these measurements the parameters and the marginal posterior distributions are estimated using the INLA method \citep{rue09}. The simulated field, the kriging prediction, and the corresponding standard errors can be seen in Figure~\ref{fig:ex2}.

\begin{figure}[t]
\begin{center}
\includegraphics[height=0.23\linewidth]{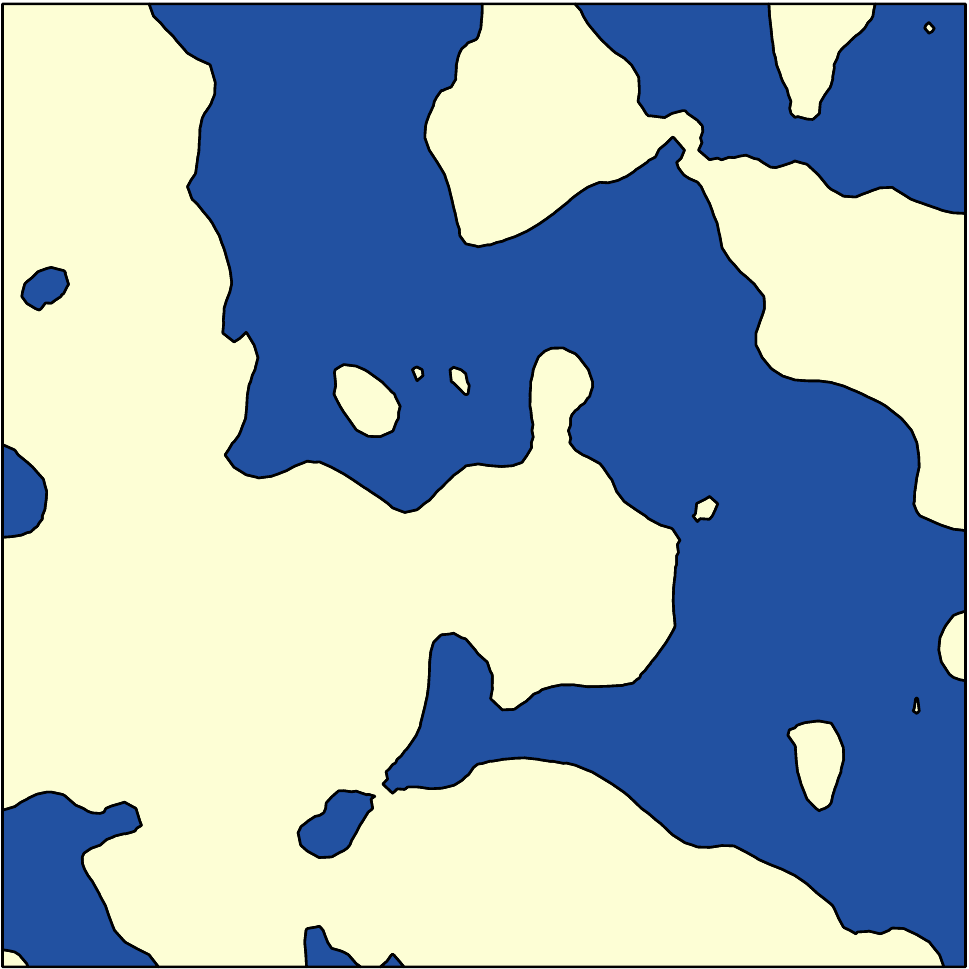}%
\hspace*{1mm}%
\includegraphics[height=0.23\linewidth]{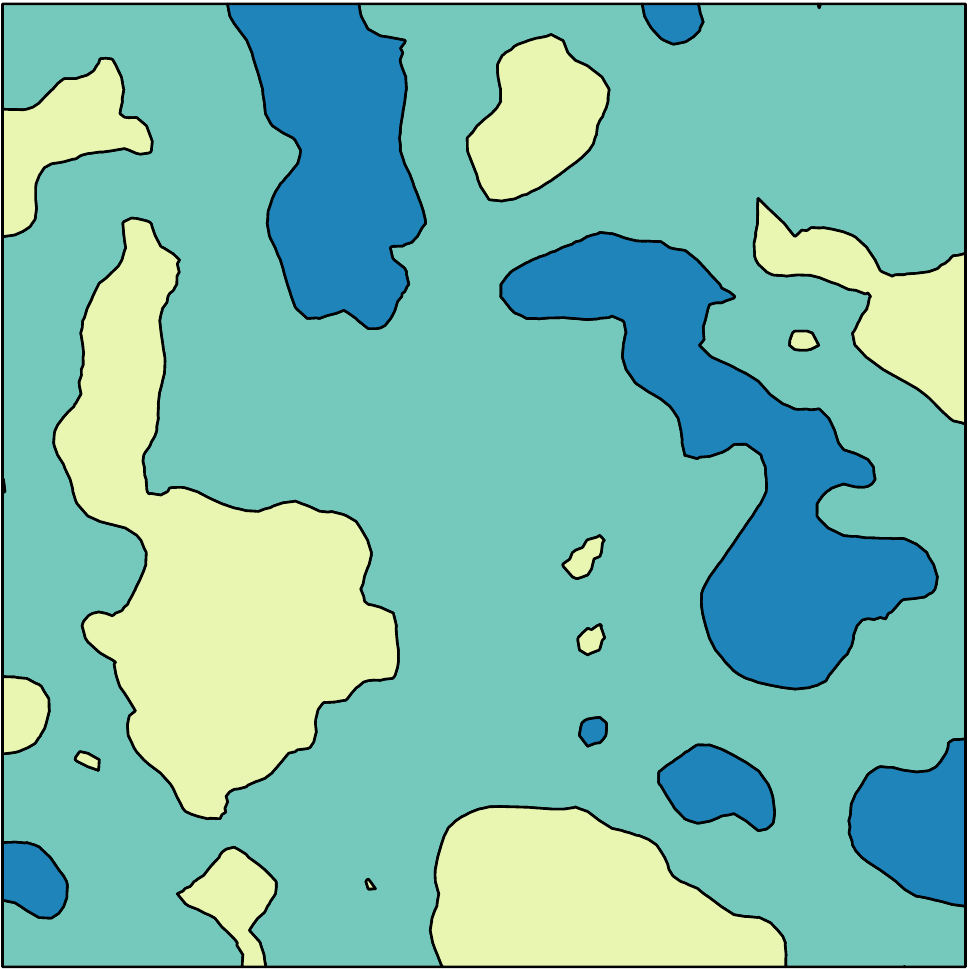}%
\hspace*{1mm}%
\includegraphics[height=0.23\linewidth]{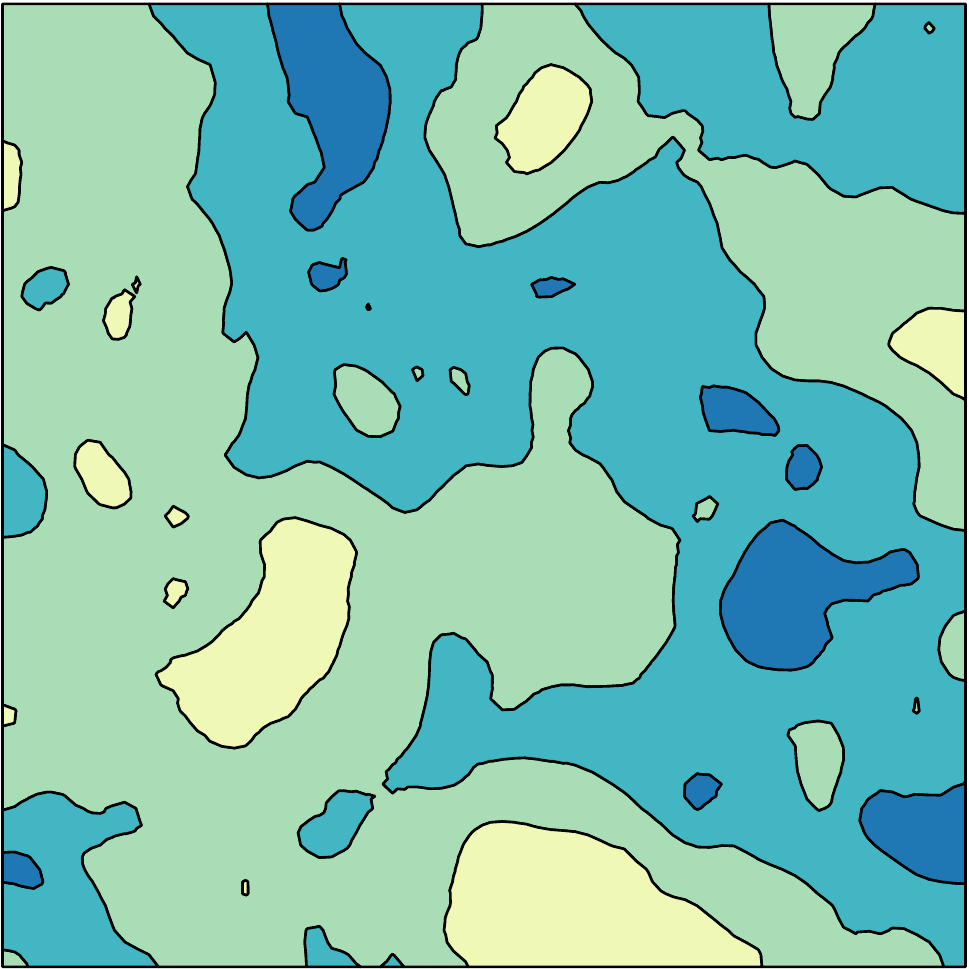}%
\hspace*{1mm}%
\includegraphics[height=0.23\linewidth]{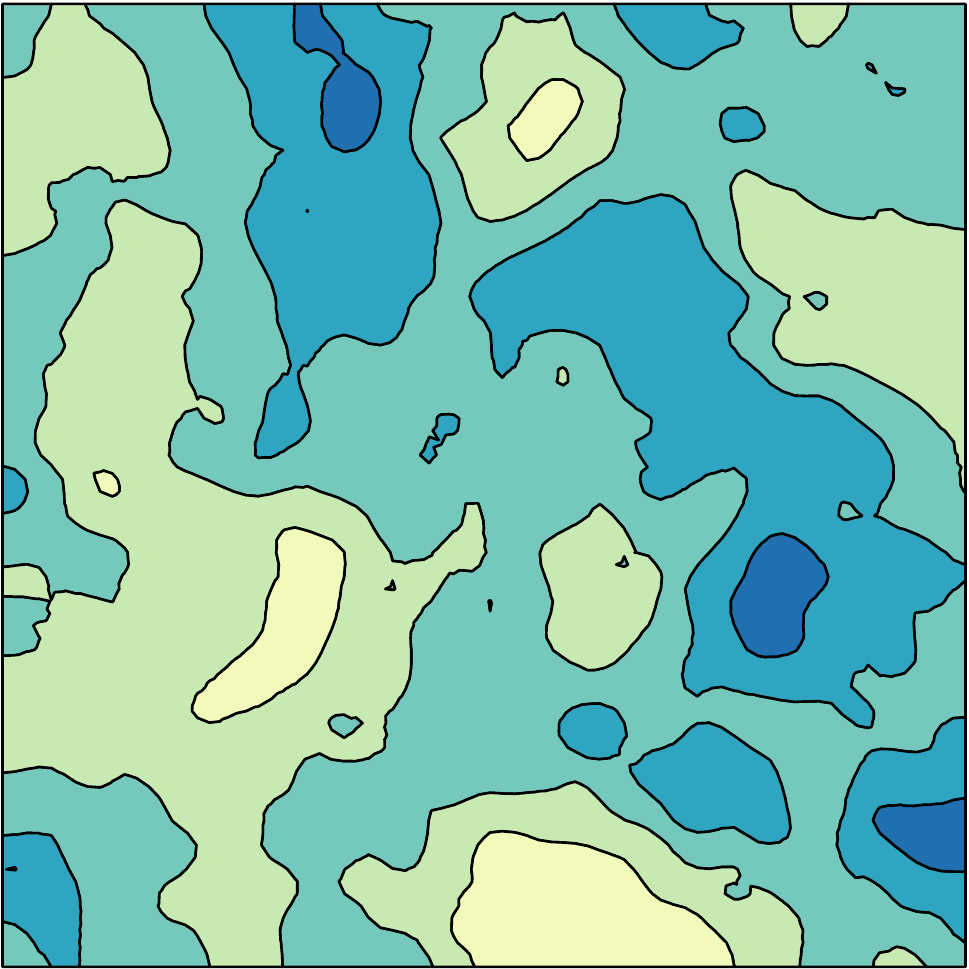}%
\hspace*{1mm}%
\includegraphics[height=0.23\linewidth]{figs/example2_mean_cb.pdf}\\[0.8mm]
\includegraphics[height=0.23\linewidth]{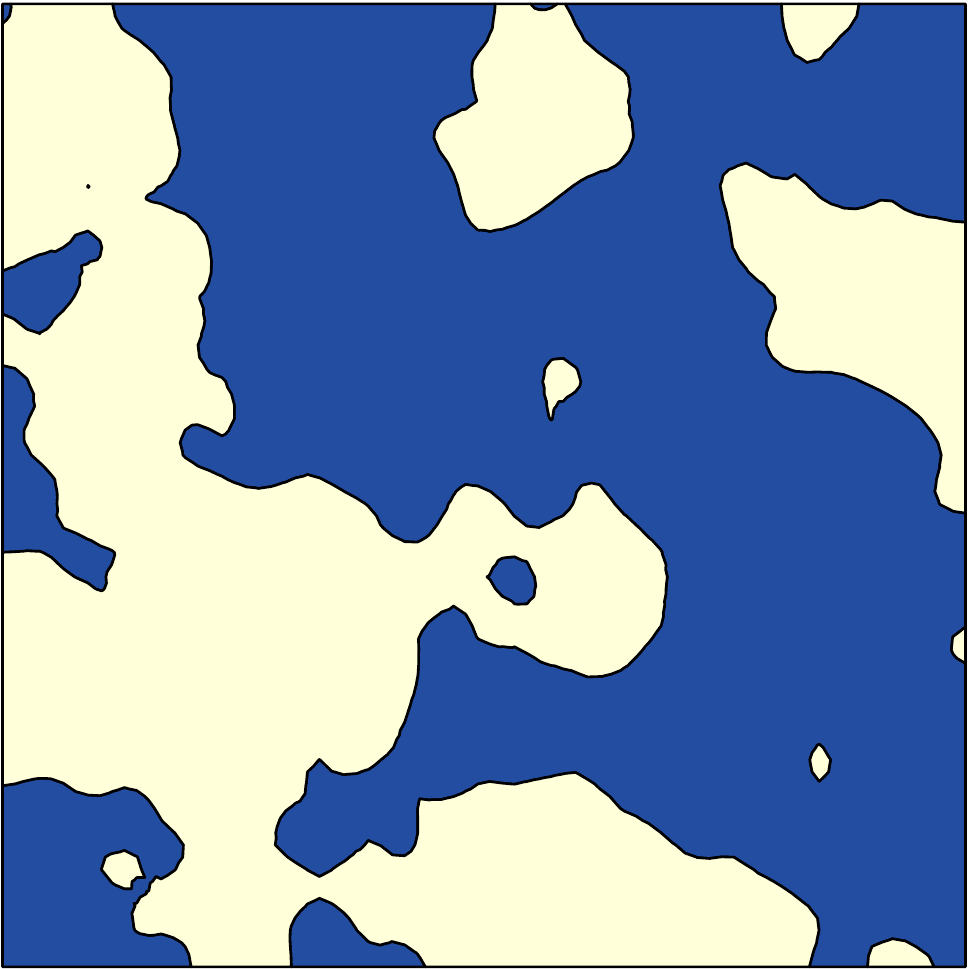}%
\hspace*{1mm}%
\includegraphics[height=0.23\linewidth]{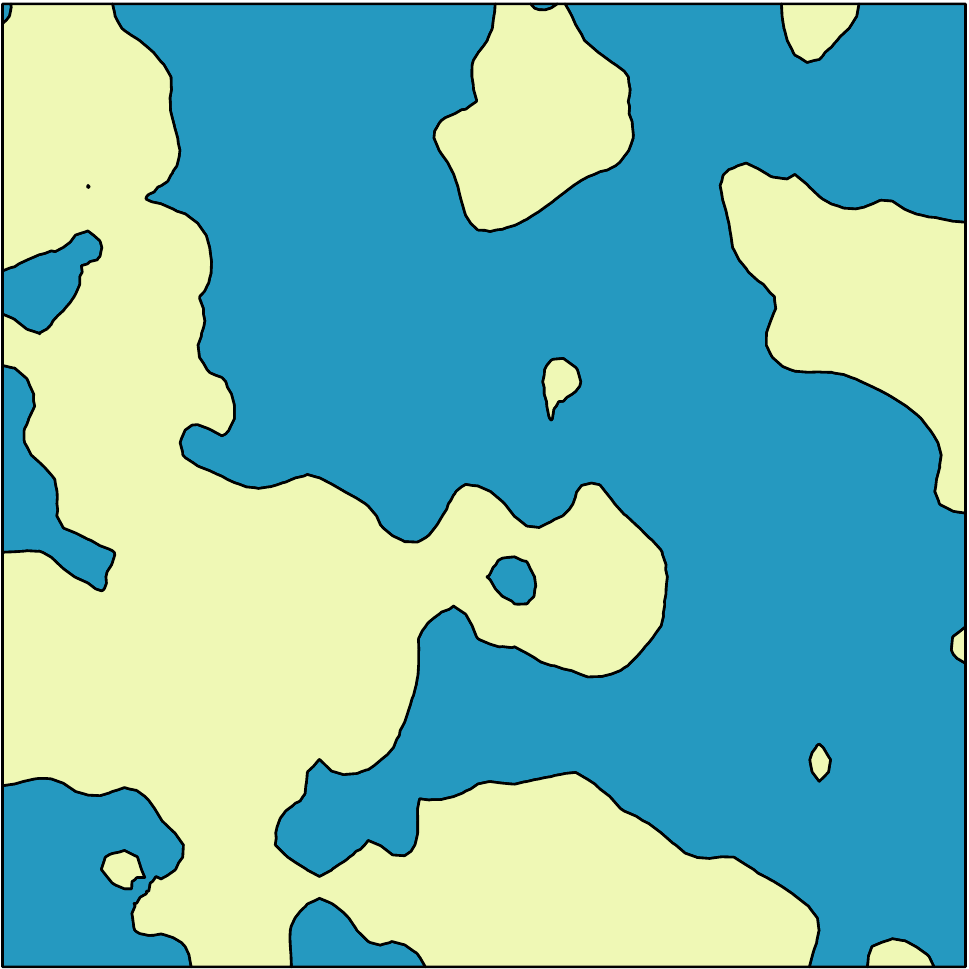}%
\hspace*{1mm}%
\includegraphics[height=0.23\linewidth]{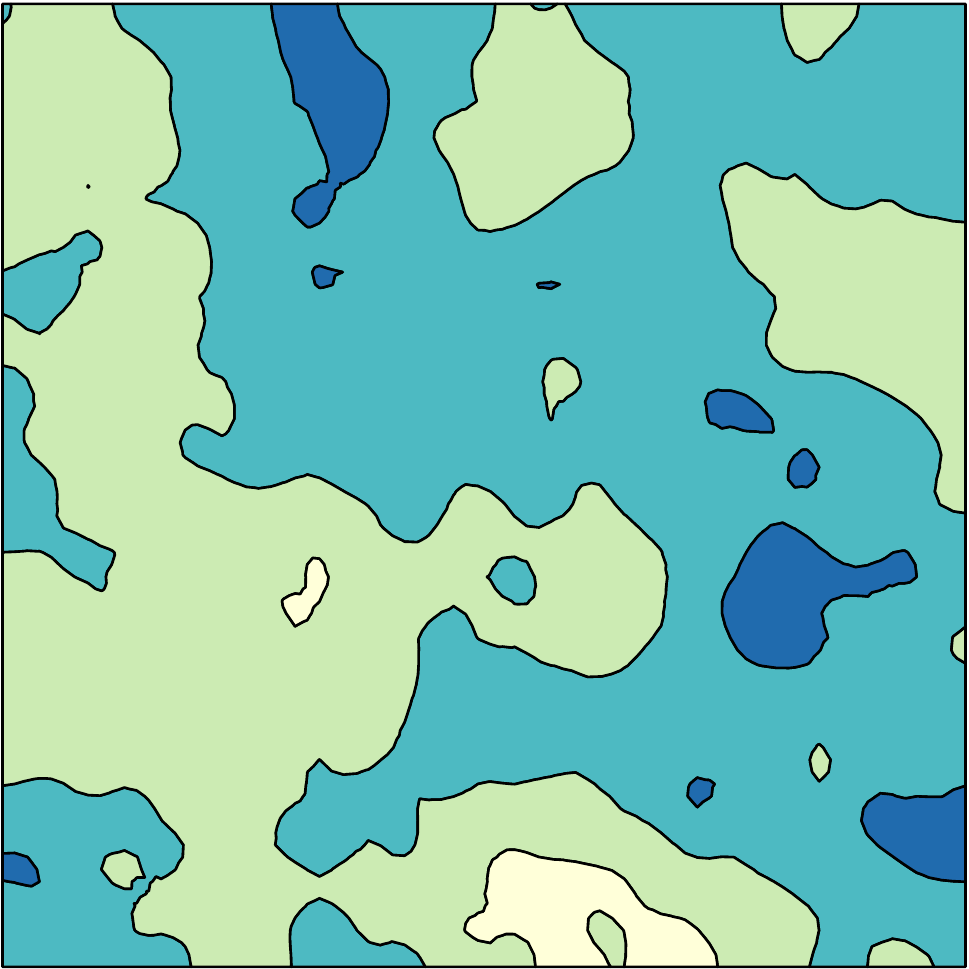}%
\hspace*{1mm}%
\includegraphics[height=0.23\linewidth]{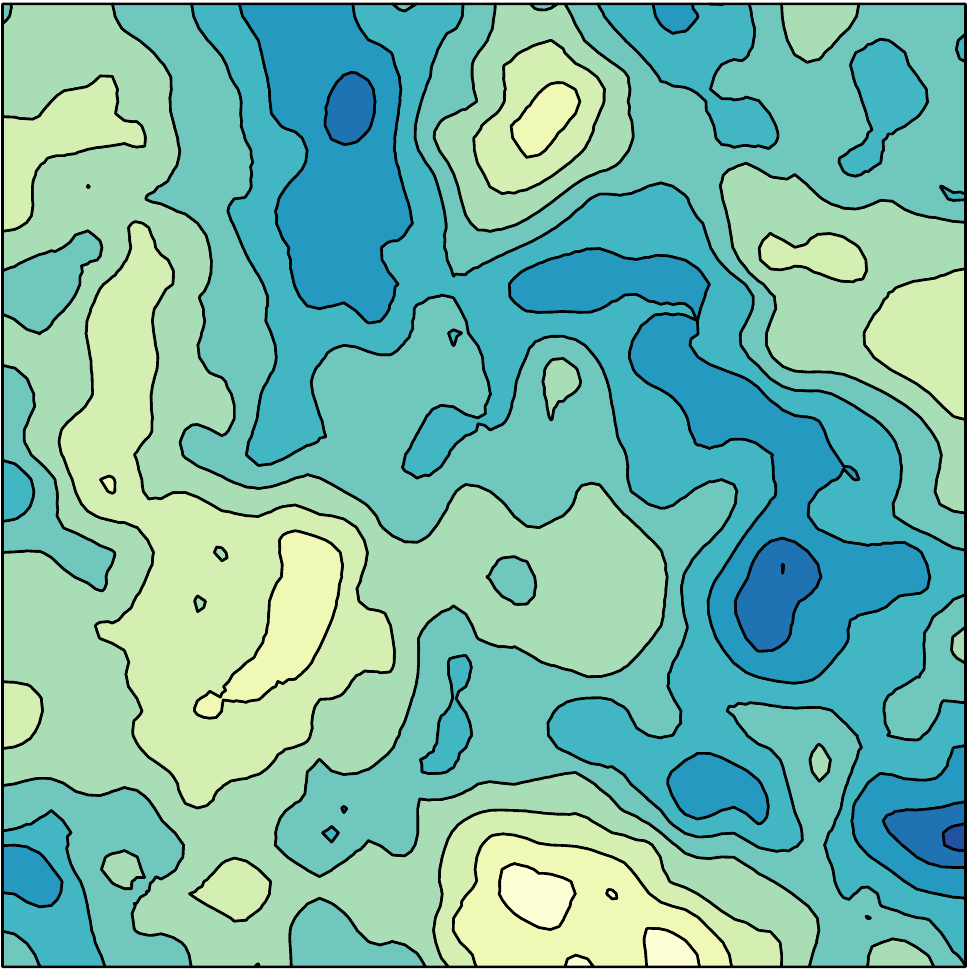}%
\hspace*{1mm}%
\includegraphics[height=0.23\linewidth]{figs/example2_mean_cb.pdf}%

\end{center}
\vspace{-0.5cm}
\caption{Standard contour maps (top) and pretty contour maps (bottom) for the simulated data example. As expected from the results in Table \ref{tab:ex2}, the second Standard contour map is similar to the third Pretty contour map.}
\label{fig:ex2_cm}
\end{figure}

The Standard contour maps, based on the kriging predictor, with $K=1,\ldots,4$ contour levels are shown in Figure~\ref{fig:ex2_cm}, as well as the first four Pretty contour maps. Table~\ref{tab:ex2} shows the $P_0$, $P_1$ and $P_2$ measures for the eight contour maps. If the $P_2$ measure should be at least $0.9$, one should among the Pretty contour maps use $K=3$, which has a level spacing of $1$. The third Pretty contour map, which has a level spacing of $0.5$, has a $P_2$ measure just below $0.9$. For the Standard maps, one should use $K=2$, which gives a level spacing that is similar to the third Pretty map.

\begin{table}[t]
\begin{center}
\begin{tabular}{lcccccccc}
\toprule
& \multicolumn{4}{c}{Standard maps} & \multicolumn{4}{c}{Pretty maps}\\
\cmidrule(r){2-5} \cmidrule(r){6-9}
K       & $1$ & $2$ & $3$ & $4$ & $3$ & $3$ & $5$ & $10$\\
Spacing & $1.577$ & $0.526$ & $0.394$ & $0.315$ &  $2$ & $1$ & $0.5$ & $0.2$\\
\cmidrule(r){2-5} \cmidrule(r){6-9}
$P_0$   & $0.613$ & $0.440$ & $0.267$ & $0.148$ & $0.616$ & $0.616$ & $0.407$ & $0.018$\\
$P_1$   & $1.000$ & $0.999$ & $0.998$ & $0.966$ & $1.000$ & $1.000$ & $0.999$ & $0.291$\\
$P_2$   & $1.000$ & $0.962$ & $0.523$ & $0.042$ & $1.000$ & $0.999$ & $0.878$ & $0.000$\\
\bottomrule
\end{tabular}
\end{center}
\caption{The quality measures $P_0$, $P_1$ and $P_2$ for the contour maps shown in Figure~\ref{fig:ex2_cm}. Here, $K$ denotes the number of contours used and Spacing refers to the spacing between the contours. If the $P_2$ measure should be at least $0.9$, one should among the Pretty contour maps use $K=3$, and for the Standard maps one should use $K=2$.}
\label{tab:ex2}
\end{table}

\begin{figure}[t]
\begin{center}

\includegraphics[height=0.23\linewidth]{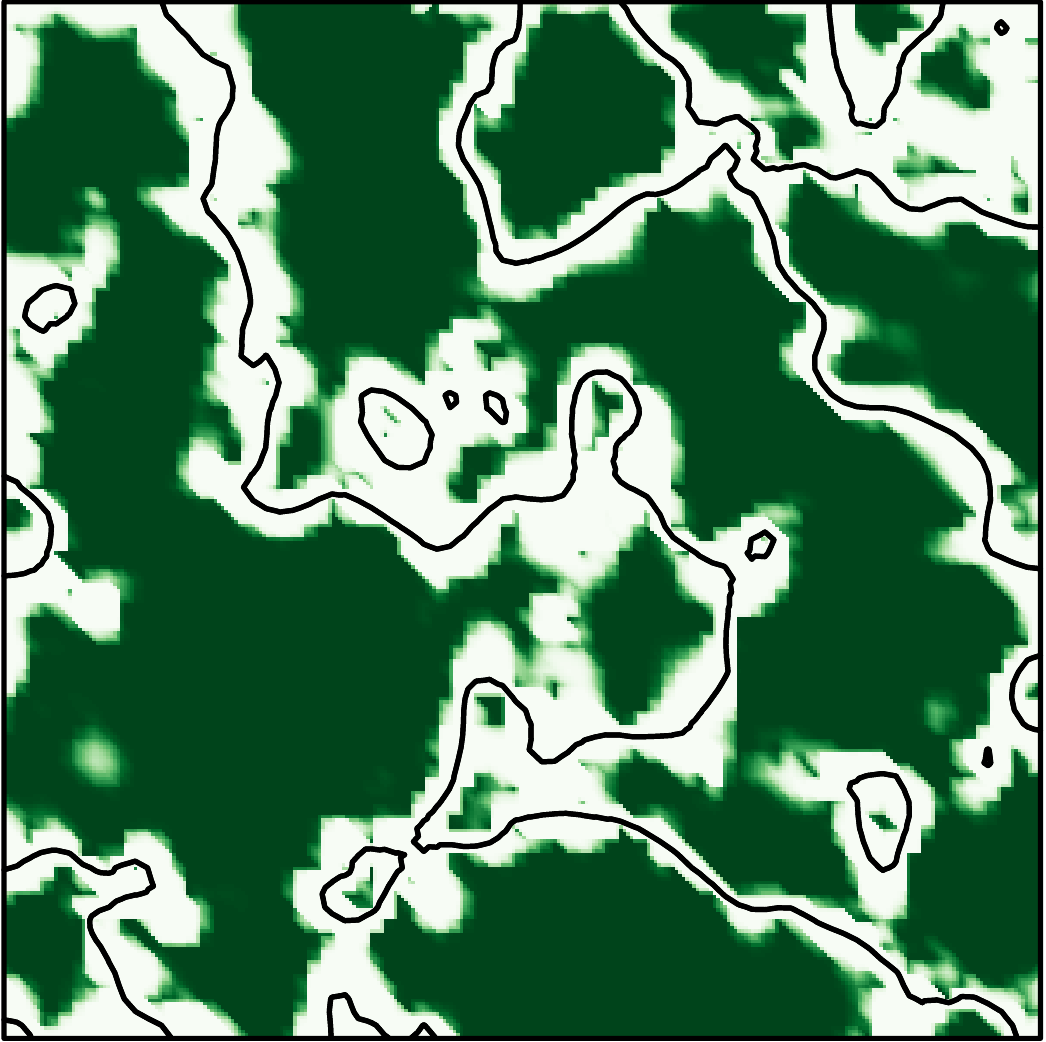}%
\hspace*{1mm}%
\includegraphics[height=0.23\linewidth]{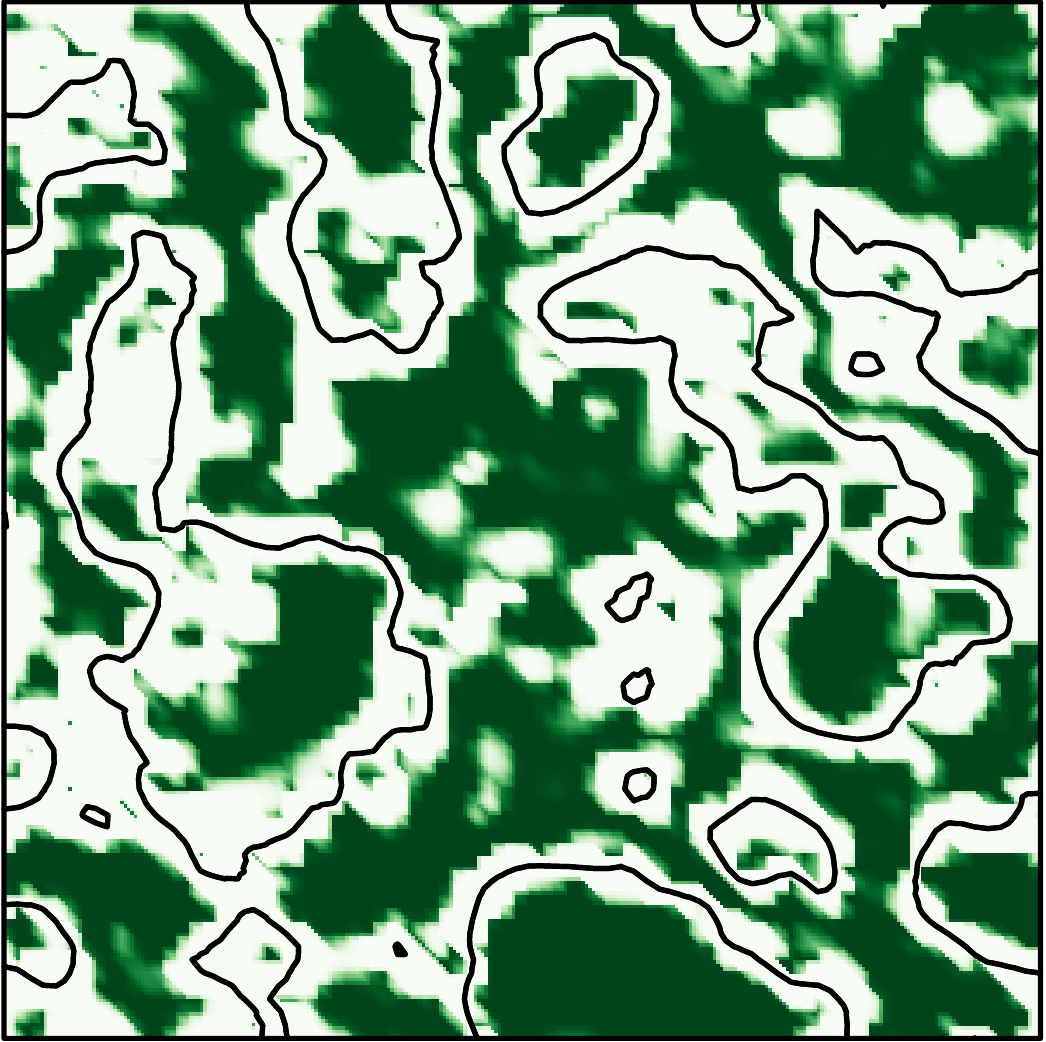}%
\hspace*{1mm}%
\includegraphics[height=0.23\linewidth]{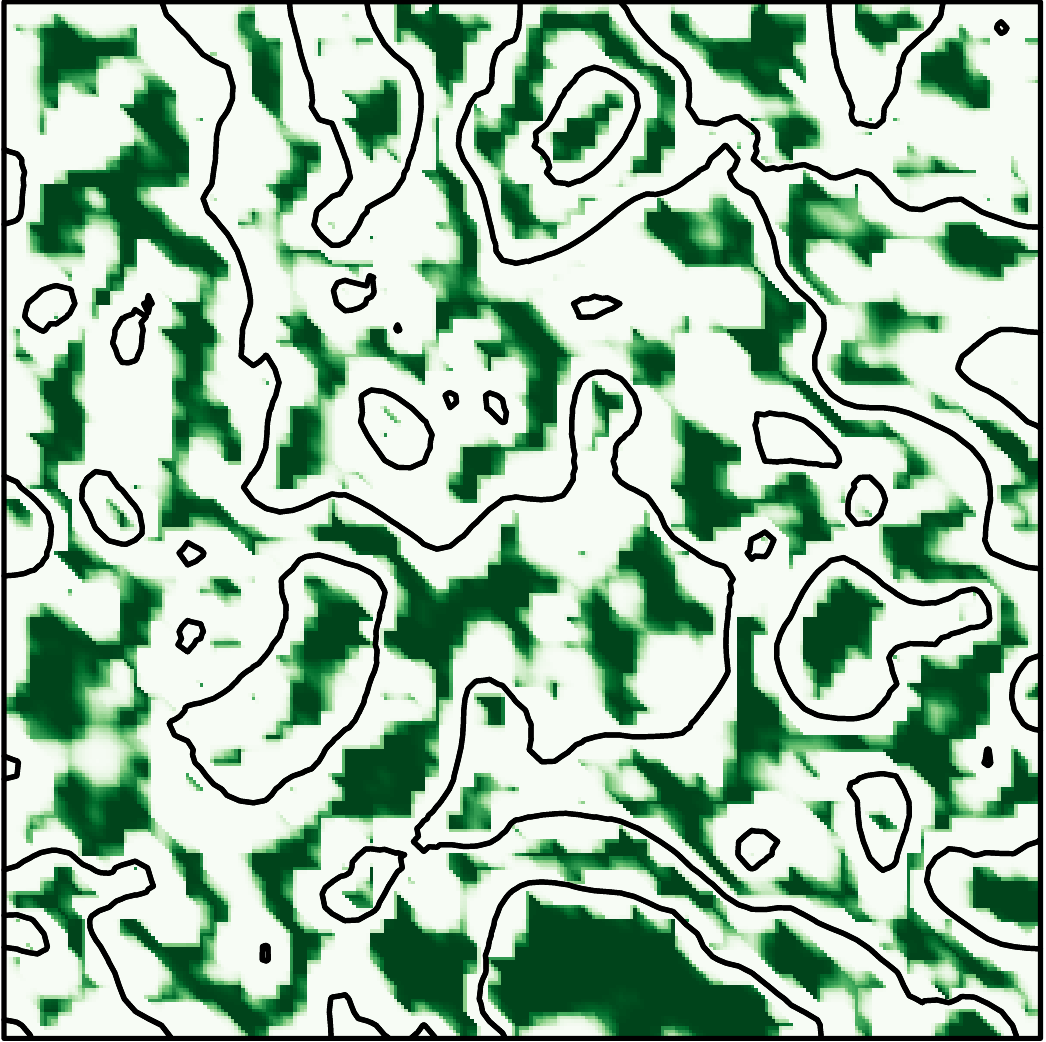}%
\hspace*{1mm}%
\includegraphics[height=0.23\linewidth]{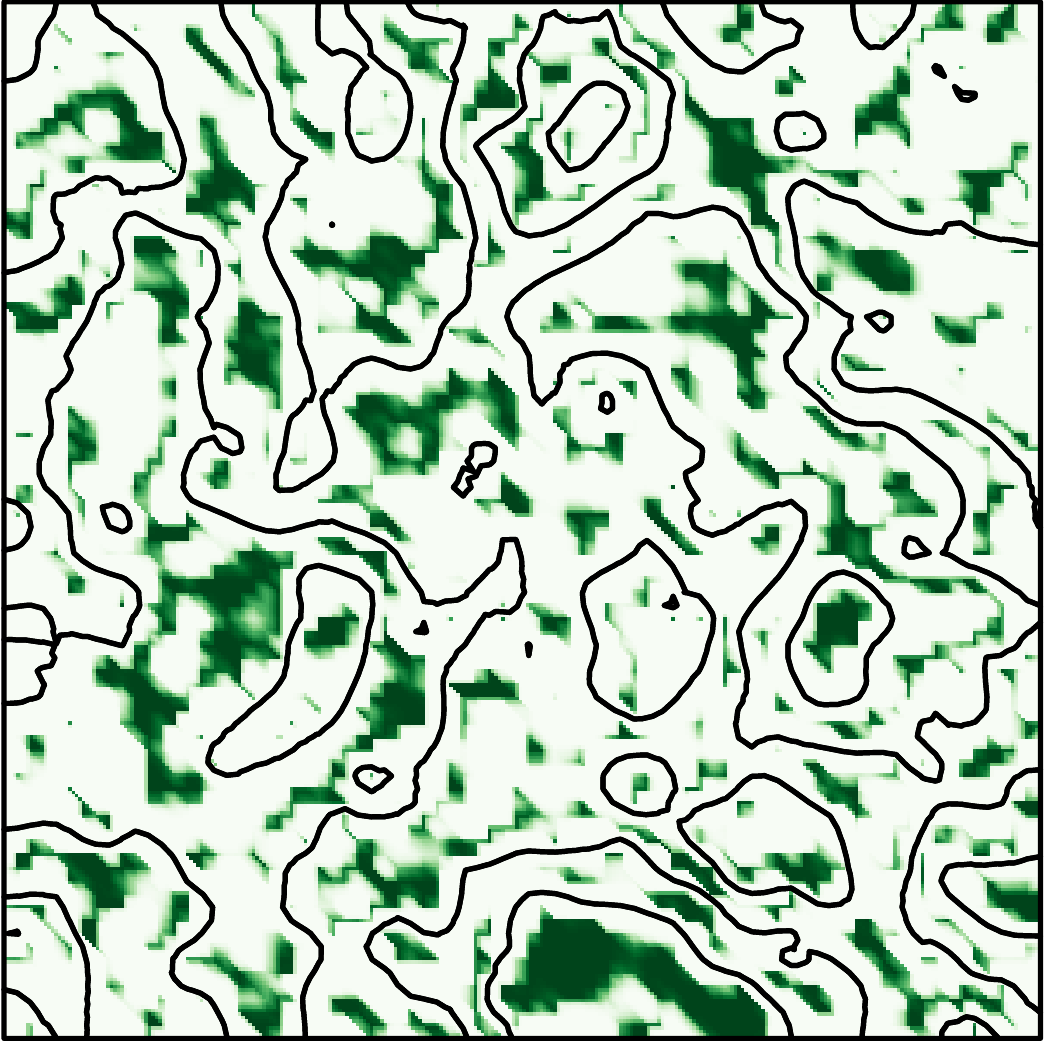}
\includegraphics[height=0.23\linewidth]{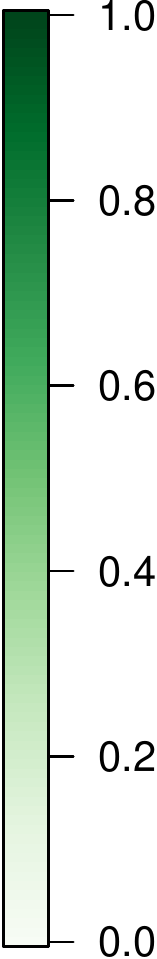}
\end{center}
\vspace{-0.4cm}
\caption{Contour map functions $F_\uu(\s)$ for the four Standard contour maps shown in Figure~\ref{fig:ex2_cm}, with the contour lines of the corresponding contour maps superimposed in black. Dark green indicates areas of confidence. The spatial average of $F_\uu(\s)$ gives the overall quality measure $P_0$, which is shown in Table \ref{tab:ex2}.}
\label{fig:ex2_F}
\end{figure}

The contour map functions $F_\uu(\s)$ for the standard contour maps can be seen in Figure~\ref{fig:ex2_F}. Recall that the contour-avoiding set $C_{\uu,\alpha}$ can be retrieved from $F_\uu(\s)$ as the set of locations where $F_\uu(\s)> 1-\alpha$. Also recall that the contour-avoiding set is the largest union of sets $M_{\uu,\alpha}^k \subseteq G_k$ so that, with probability $1-\alpha$, the field satisfies $u_k < x(\s) < u_{k+1}$ for $\s\in M_{\uu,\alpha}^k$. The $P_0$ measures corresponding to the contour map functions shown in Figure~\ref{fig:ex2_F} are shown in Table \ref{tab:ex2}. 

\section{An application to temperature visualization}\label{sec:applications}
\begin{figure}[t]
\begin{center}
\resizebox{\linewidth}{!}{\includegraphics{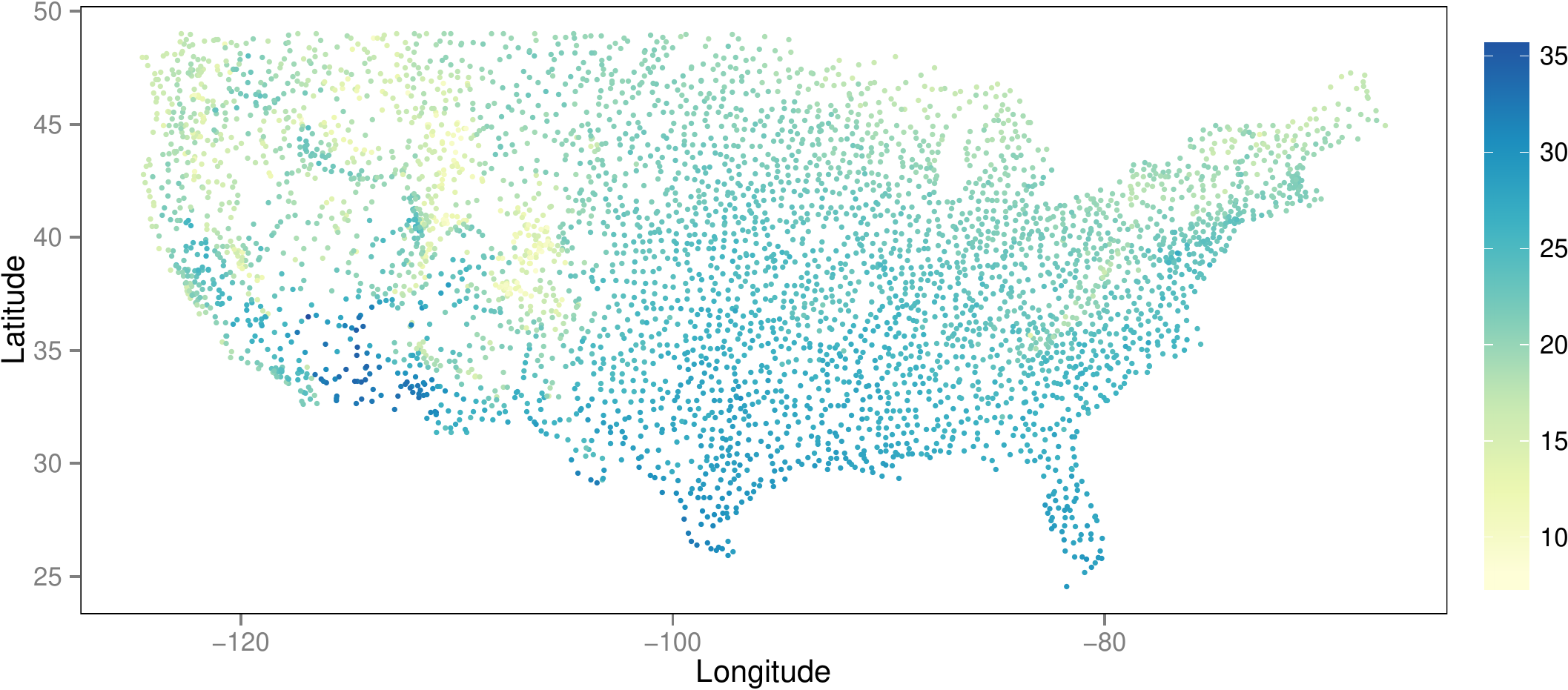}}%
\end{center}
\vspace{-0.5cm}
\caption{Mean summer temperature measurements for 1997. The spatial observation pattern is fairly uniform with the exception of the mountainous region in the west.}
\label{fig:temp_data}
\end{figure}

In this section, we illustrate the contour map methods by looking at the problem of estimating the mean summer temperature for the US. The data we use is available at \texttt{www.image.ucar.edu/Data/US.monthly.met} and was created from the data archives of the National Climatic Data Center. The data set contains daily measurements of min and max temperatures at approximately 8000 stations in the United States with a temporal coverage from 1895 to 1997. As a simple example, we focus on the mean summer temperature (June - August) for the year 1997. For each station, the mean temperature is defined as the mean of the recorded max and min temperatures, and the values are averaged over the summer months to obtain the mean summer temperature. The resulting data is shown in Figure~\ref{fig:temp_data}.

As a first model, we assume that the temperature measurements are observations of the true temperature field, $x(\cdot)$, under mean-zero Gaussian measurement noise, $y_i = x(\s_i) + \epsilon_i$, where $\epsilon_i$ are iid $\pN(0,\sigma^2)$ variables. The latent field is modeled as $x(\s) = \beta_0 + \xi(\s)$ where $\beta_0$ is the mean temperature and $\xi(\s)$ is a mean-zero Gaussian Mat\'ern field as in the example in the previous section. We estimate the model parameters using INLA and compute the kriging predictor of $x(\cdot)$ on a regular grid in the region.

\begin{figure}[t]
\begin{center}
\resizebox{0.45\linewidth}{!}{\includegraphics{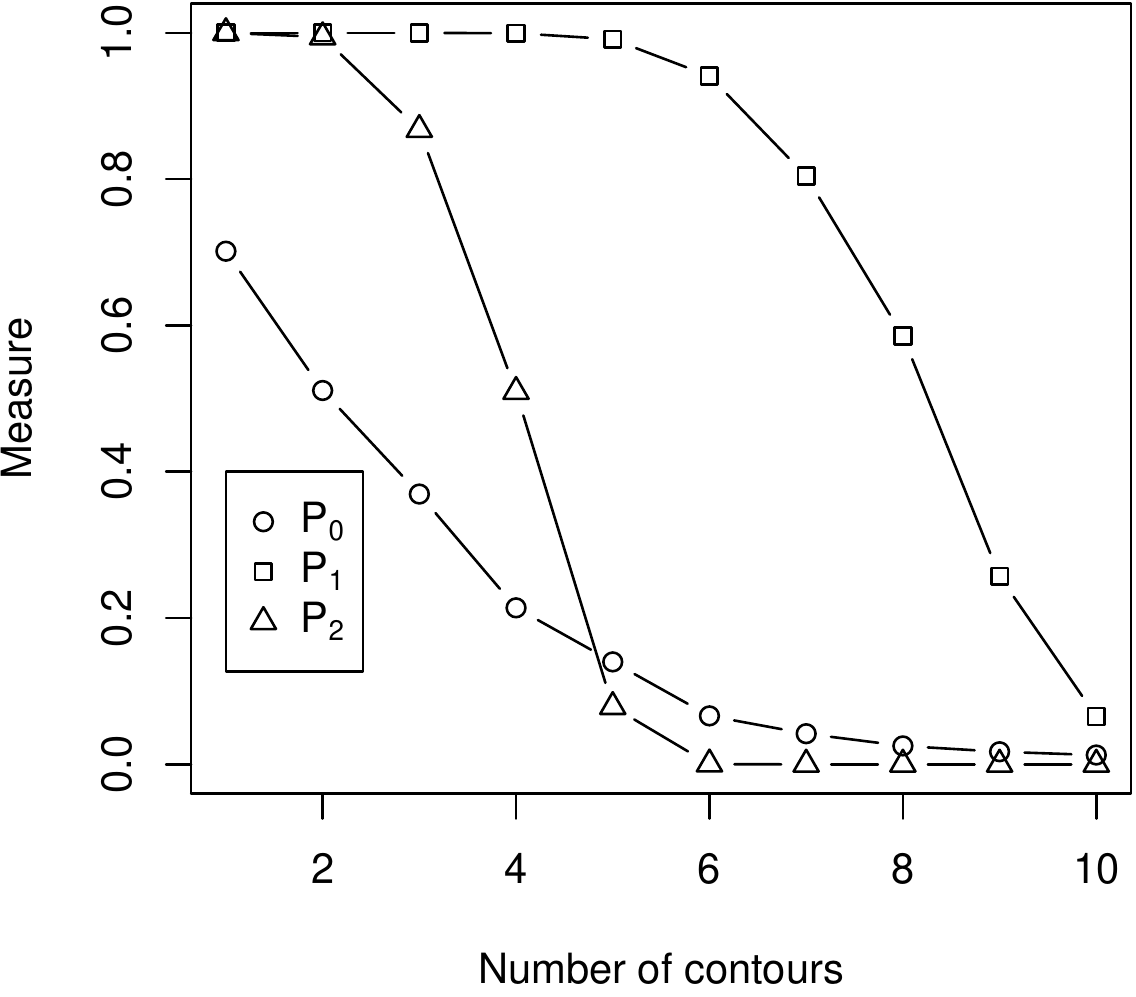}}
\resizebox{0.45\linewidth}{!}{\includegraphics{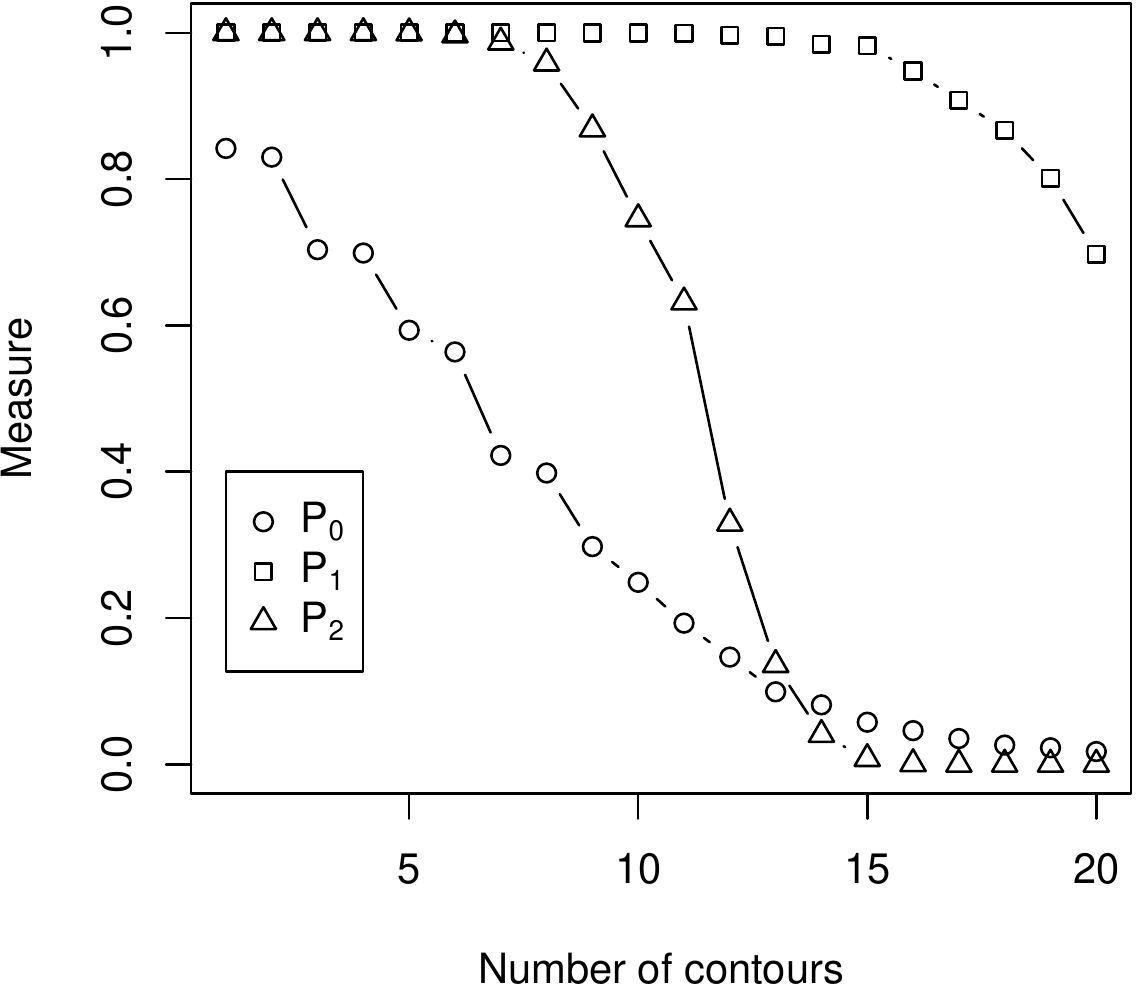}}
\end{center}
\vspace{-0.5cm}
\caption{Contour map quality measures for the temperature application using a
  model without covariates (left panel) and using a model with
  altitude as a covariate for the mean (right panel). Squares indicate
  the $P_1$ measure, triangles the $P_2$ measure, and circles the
  $P_0$ measure. As expected from the definition, the $P_1$ measure admits approximately twice as many contour levels as $P_2$ for a given target probability. The $P_0$ measure has a different type of behavior, and the associated spatial pattern in Figure \ref{fig:temp_F} is perhaps more useful.}
\label{fig:measures}
\end{figure}

We calculate contour maps with $K=1,\ldots,10$ levels  based on the kriging predictor for the estimated model. The $P$-measures for the contour maps are shown in the left panel of Figure~\ref{fig:measures}, and one can see that we are only allowed to use $2$ contours if we want a contour map with $P_2 \geq 0.9$.

However, modeling the temperature as a stationary Mat\'ern field is too simplistic. Therefore, we add altitude as a covariate for the mean of the field. Thus, the second model has $x(\s) = \beta_0 +  a(\s)\beta_1 + \xi(\s)$ where $a(s)$ is the altitude covariate. The altitude is available for each measurement location, and altitude data for the grid to which we do predictions was obtained from the ETOPO1 1 Arc-Minute Global Relief Model \citep{ETOPO1}.

\begin{figure}[t]
\begin{center}
\begin{minipage}[b]{0.49\linewidth}
\centering
\includegraphics[width=0.95\linewidth]{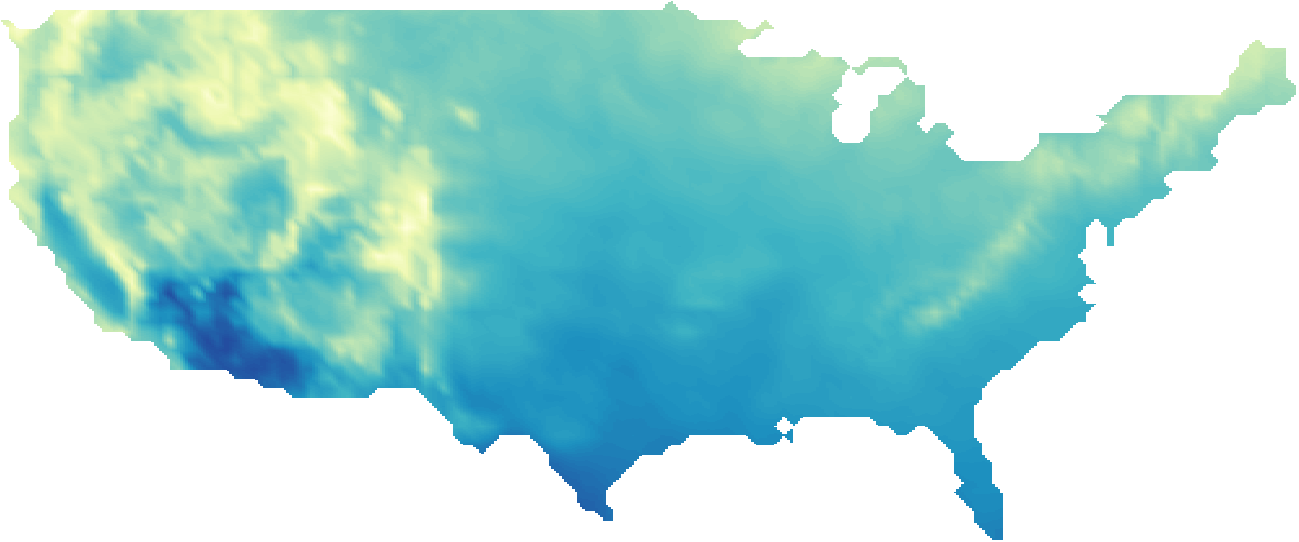}%
\hspace*{-0.5cm}\includegraphics[width=0.075\linewidth]{figs/temp_mean_cbar.pdf}%
\end{minipage}
\begin{minipage}[b]{0.49\linewidth}
\centering
\includegraphics[width=0.95\linewidth]{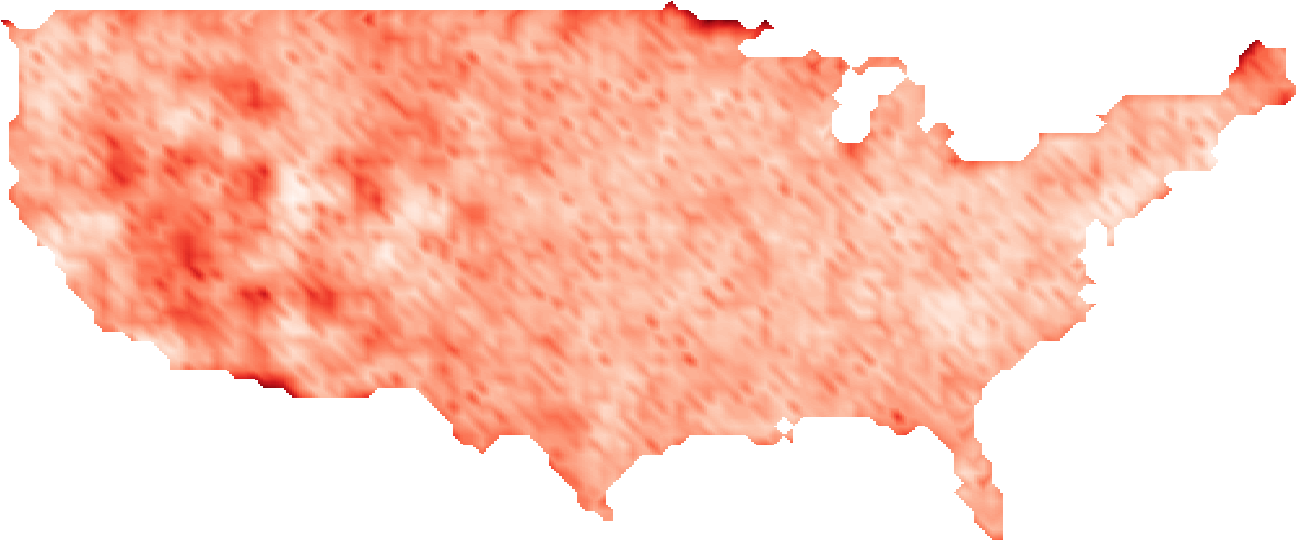}%
\hspace*{-0.5cm}\includegraphics[width=0.075\linewidth]{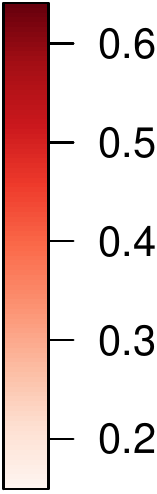}%
\end{minipage}
\end{center}
\caption{Posterior mean (left) and posterior standard deviations (right) for the temperature application using a model with altitude as a covariate for the mean. The spatial posterior mean pattern is dominated by the covariate, and the standard deviation pattern matches the spatial observation density seen in Figure \ref{fig:temp_data}.}
\label{fig:temp_mean}
\end{figure}

\begin{figure}[t]
\begin{center}
\begin{minipage}[b]{0.49\linewidth}
\centering
\includegraphics[width=0.95\linewidth]{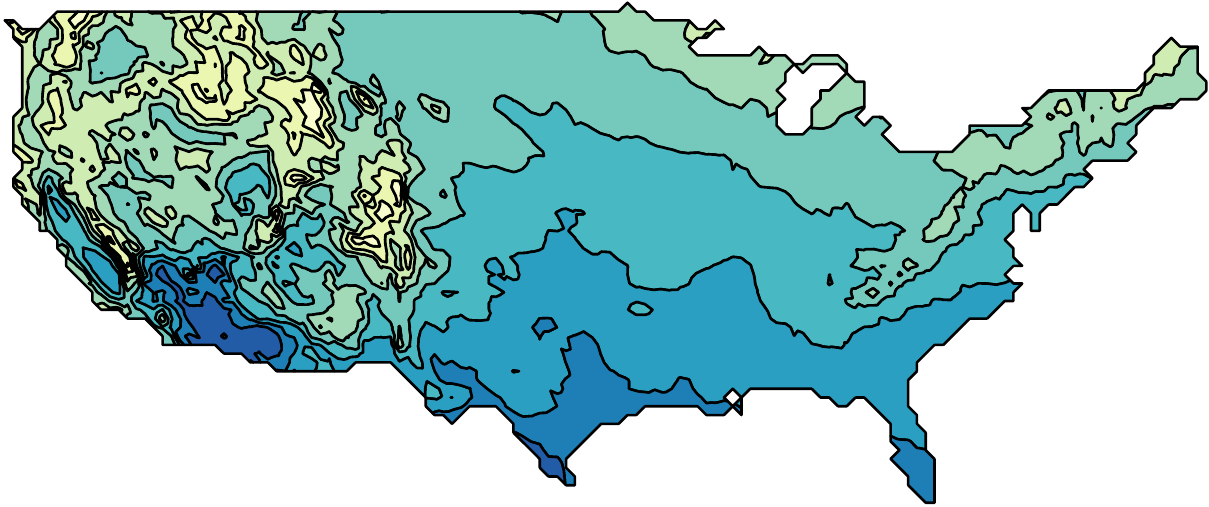}
\hspace*{-0.5cm}\includegraphics[width=0.075\linewidth]{figs/temp_mean_cbar.pdf}
\end{minipage}
\begin{minipage}[b]{0.49\linewidth}
\centering
\includegraphics[width=0.95\linewidth]{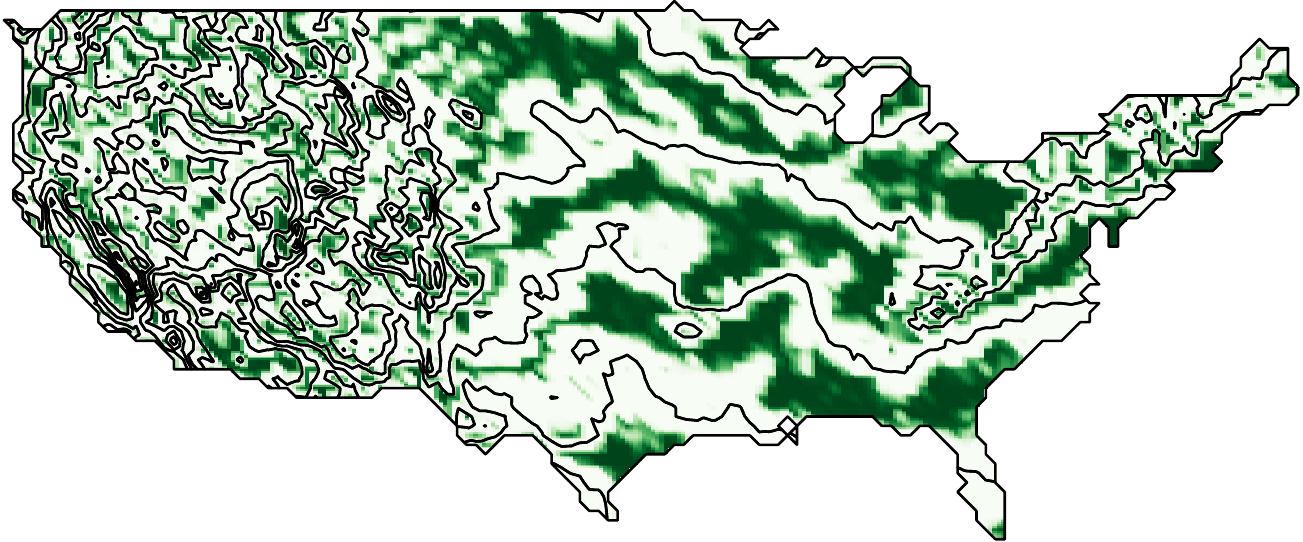}
\hspace*{-0.5cm}\includegraphics[width=0.075\linewidth]{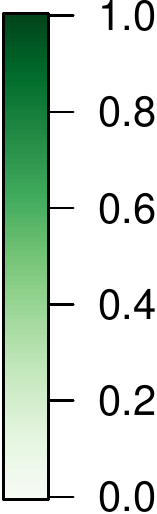}
\end{minipage}
\end{center}
\caption{A contour map with eight contours with $P_2 = 0.958$ (left) and the corresponding contour map function $F_\uu(\s)$ with spatial average $P_0 = 0.206$ (right). As expected, the confidence is highest in the large flat regions in the central area due to the clear latitude effect. }
\label{fig:temp_F}
\end{figure}

The kriging predictor and standard errors using the second model can be seen in Figure~\ref{fig:temp_mean}. We again calculate contour maps and quality measures, now for $K=1,\ldots,20$ levels. The resulting quality measures can be seen in the right panel of Figure~\ref{fig:measures}. For this model, we are allowed to draw at most eight contours if we want a contour map with $P_2 \geq 0.9$. Figure~\ref{fig:temp_F} shows the Standard contour map with $10$ levels and the corresponding contour map function. For this contour map, we have $P_2=0.958$ and $P_0=0.206$.

\section{Discussion}\label{sec:conclusions}
Although contour maps are widely used, relatively little research has focused on quantifying their statistical properties. We have here
defined three contour map quality measures inspired by \citet{Polfeldt99} that can be used to assess how appropriate a contour map is a for a given problem. The $P_0$ measure is based on the spatially interpretable contour map function $F_\uu(\cdot)$ and can be used as a measure of the proportion of the domain in which the field lies safely between the contour levels.  For choosing an appropriate number of contour lines, or an appropriate level spacing, $P_2$ appears to be the most useful measure, as it has a precise definition that is also practically interpretable.  It gives the probability that the level crossing of the in-between levels associated with each $G_k$ all fall inside their respective $G_k$ region, making $(G_1,\dots,G_K)$ a joint credible region collection for those level crossings, as illustrated in
Figure~\ref{fig:illustration2}(right).  The $P_1$ measure is similar in spirit to $P_2$, but with no obvious advantages.

An intuitively interpretable approach to choose the number of equally spaced levels is to find $K$ such that $P_2$ is above some threshold.  For a joint credibility of 90\%, say, choose the largest $K$ or the smallest spacing such that $P_2 \geq 0.9$.  For a more permissive choice, require $P_2 \geq 0.5$ instead. In our experience, $P_2$ transitions quickly from $1$ to $0$, so the different thresholds are unlikely to give very different results.

The computational methods are constructed on a discrete space, and the
results have to be interpreted appropriately.  In particular, we
briefly presented methods to be used for continuous spatial domains,
based on interpolation of the contour map function $F_\uu(s)$. We also
showed that if the computational analysis is carried out on the same
spatial scale as the true process, the credible contour regions
resulting from the interpolation methods have the desired coverage
probability.  Further, when the true field has a fine scale structure,
the analysis has to be carried out on a high resolution in order to
capture small contour curve segments.  We provided simple \emph{ad hoc} rules for choosing the computational scale, but further studies are required to give more precise guidance. It should also be noted that the computational scale is of importance not only for the contour maps, see \cite{bolin09b} for a more detailed discussion of this issue.

Efficient computational methods for finding the required quantities
use the methods by \cite{bolin12}. They work for the class of latent
Gaussian models and are especially suitable for latent Gaussian Markov
random field models. All methods discussed here are implemented in the
R package \texttt{excursions}.

\section*{Acknowledgements}
The authors are grateful to the reviewers and editors for their
helpful and constructive comments, that served to greatly improve the manuscript, and wish to acknowledge the importance of Prof Peter Guttorp, who has been a strong champion of the topic. The first author has been supported by the Knut and Alice Wallenberg foundation.